\documentclass[a4paper, accepted=2026-07-01]{quantumarticle}
\pdfoutput=1
\usepackage{hyperref}
\usepackage{nameref}
\usepackage{cite}
\usepackage{graphicx,color}
\usepackage{bm}
\usepackage{dsfont,amsmath,amssymb}
\usepackage{amsthm}

\newcommand{\bra}[1]{\mbox{$\langle #1 |$}}
\newcommand{\ket}[1]{\mbox{$| #1 \rangle$}}

\newcommand{\ketbra}[2]{\ket{#1}\!\bra{#2}}

\newcommand{\Tr}{\mathrm{Tr}}

\newcommand{\ee}{\mathrm{e}}
\newcommand{\ii}{\mathrm{i}}

\begin{document}

\title{From Superradiance to Superabsorption: An Exact Treatment of Non-Markovian Cooperative Radiation}
	\author{Ignacio González}
    \orcid{0009-0000-8199-6935}
	\email{ignago10@ucm.es}
	\affiliation{Departamento de F\'{\i}sica Te\'orica, Facultad de Ciencias F\'isicas, Universidad Complutense, 28040 Madrid, Spain.}
	\author{\'Angel Rivas}
    \orcid{0000-0002-0636-2446}
	\email{anrivas@ucm.es}
	\affiliation{Departamento de F\'{\i}sica Te\'orica, Facultad de Ciencias F\'isicas, Universidad Complutense, 28040 Madrid, Spain.}
	\affiliation{CCS-Center for Computational Simulation, Campus de Montegancedo UPM, 28660 Boadilla del Monte, Madrid, Spain.}

%\date{\today}

\begin{abstract}
We investigate the emergence of cooperative radiation phenomena in ensembles of two-level atoms coupled to a lossy resonant cavity beyond the Markovian and mean-field approximations. By deriving a complete analytical solution for the two-emitter case and employing a numerically exact method for larger ensembles, we characterize the full transition from Markovian to non-Markovian collective dynamics for systems of up to $10^3$ emitters. Our results reveal three distinct regimes: a Markovian phase exhibiting the standard superradiant burst, a non-Markovian phase featuring spontaneous superabsorption of the emitted field, and a critical regime marked by pulsed collective emission. We show that the critical spectral width separating these behaviors increases monotonically with the number of emitters, demonstrating that environmental memory effects can be enhanced by cooperativity. Finally, we find that the superradiant scaling of the peak intensity progressively degrades with increasing system size, approaching a subquadratic law in the limit of a perfect cavity. In this regime, spontaneous superabsorption emerges as a distinct manifestation of non-Markovian cooperativity.
\end{abstract}

\maketitle

\section{Introduction} Since Dicke’s seminal prediction that an ensemble of closely spaced atoms can radiate cooperatively with an intensity far exceeding the sum of their independent emissions \cite{Dicke}, the phenomenon of superradiance has stood as a cornerstone example of collective quantum behavior. A hallmark of this cooperative emission is the appearance of a sharply peaked intensity burst at a finite delay time, with a maximum that scales quadratically with the number of atoms \cite{Dicke,Agarwal,Rehler,Bonifacio71a,Bonifacio71b,Eberly,GrossHaroche,MaWo95,BrPe02}. This striking prediction has since been confirmed in a variety of experimental platforms \cite{Skribanowitz, GrossHarocheExp,Pavolini,DeVoe,Scheibner,Mlynek,Solano,Kim,Chen,Asenjo23}. The canonical theoretical description of superradiance, however, relies on mean-field and Markovian approximations. These overlook crucial dynamics that emerge when the assumption of a memoryless reservoir breaks down, giving rise to non-Markovian effects in which information or energy flows back from the environment to the emitters. Understanding how cooperative emission unfolds in the presence of such memory effects is thus essential for a complete picture of superradiance.

Broadly, memory effects in these settings can arise from two distinct physical origins. The first is due to retardation, where the finite travel time of light between distant emitters induces memory, even in environments with a flat spectral density like one-dimensional waveguides. This leads to phenomena such as modified decay rates and the formation of atom-photon bound states \cite{ Dinc1, Dinc2, Sinha, Qiu, Capurso25}. The second, and perhaps more common, origin is the presence of a structured photonic environment. When emitters are coupled to systems with a strongly frequency-dependent spectral density, such as photonic band-gap materials or resonant cavities, the environmental memory reshapes the collective emission, leading to effects ranging from the suppression of decay to the emergence of novel scaling laws \cite{John,  Vats, TudelaCirac1, TudelaCirac2, Thanopoulos, Hou}. While much of this work has focused on emission dynamics, the static ground-state properties and phase diagram of such systems have also been studied in the ultrastrong-coupling regime \cite{DeBernardis}. Notably, a full understanding of non-Markovian superradiance remains an open question, since most theoretical treatments are constrained to specific regimes, such as single excitation states, small ensembles or short times.

In addition, recent work has also turned attention to the inverse process, dubbed as superabsorption, in which absorption is collectively enhanced. Besides its conceptual interest as the absorption counterpart of superradiance, it holds promise for quantum technologies such as fast-charging quantum batteries and efficient energy-harvesting devices \cite{Quach2022, Kamimura22}. While experiments and theoretical proposals have demonstrated that superabsorption can be engineered through external control fields \cite{Higgins14, Yang21}, it remains unclear whether such enhancement can emerge spontaneously from intrinsic system–reservoir dynamics. 

Answering these questions requires obtaining the exact dynamics of a large number of emitters, a task which is notoriously difficult. In this work, we advance in this direction by providing comprehensive and nonperturbative solutions to the dynamics of initially excited two-level atoms coupled to a lossy resonant cavity. Specifically, after introducing the physical model and reviewing the well-known Markovian results in Section II, we derive in Section III analytical results for the exact non-Markovian dynamics of small ensembles, obtaining a complete solution for the two-body case,  $N=2$. This constitutes a particularly valuable result given the scarcity of analytical solutions for genuinely non-Markovian dynamics, especially beyond the single-emitter case. To deal with ensembles of increasing size, in Section IV we combine the pseudomode method with a weak symmetry of the resulting master equation. Using this technique, we solve the dynamics of considerably large systems in a numerically exact way and identify three distinct dynamical regimes: a Markovian one exhibiting the familiar superradiant burst; a non-Markovian regime marked by spontaneous superabsorption following initial emission; and a critical case characterized by collective pulsed emission. We further investigate how the peak radiated intensity scales with both the reservoir spectral width and the number of emitters, and show that the maximum reabsorbed intensity mirrors its emissive counterpart, displaying a superlinear scaling with $N$ induced by collective effects.

Altogether, our work highlights how reservoir memory fundamentally reshapes collective light–matter dynamics, uncovering a novel emission–absorption phenomenology absent from standard Markovian descriptions.

\section{Physical Model}

\begin{figure}[t]
	\includegraphics[width=\columnwidth]{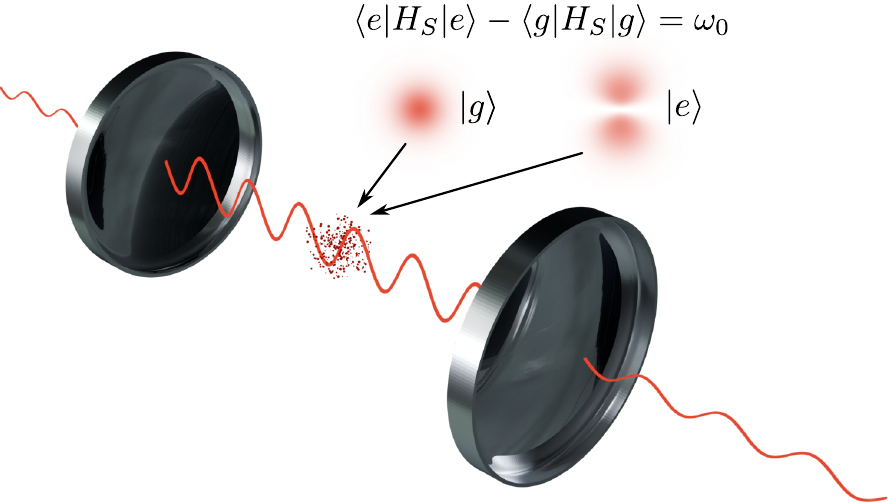}
	\caption{Visual representation of the physical model. $N$ identical two-level atoms interact with a resonant QED cavity mode with losses modeled by the spectral density \eqref{spectraldensity}. The energy gap between the ground state $\ket{g}$ and the excited state $\ket{e}$ is given by $\omega_0$ (units of $\hbar=1$). The atoms are assumed to be close enough to experience the same phase of the electromagnetic field, but not so close as to interact significantly with each other. Orbital shapes are used for illustrative purposes only.}
	\label{Fig:0}
\end{figure}

As illustrated in Fig. \ref{Fig:0}, we consider an ensemble of $N$ two-level independent atoms with free Hamiltonian $H_S=\frac{\omega_0}{2}\sum_{n=1}^{N}\sigma_z^{(n)}=\omega_0 J_{z}$, with $J_z=\frac{1}{2}\sum_{n=1}^{N}\sigma_z^{(n)}$, inside a resonant QED cavity modeled by the Hamiltonian $H_B=\sum_{\bm{k}} \omega_k a^\dagger_{\bm{k}}a_{\bm{k}}$, with $\omega_k=|\bm{k}|$ and $a_{\bm{k}}$ the annihilation operator for a photon with wavevector $\bm{k}$. We write the ground and excited states of the $n$th atom as $\ket{g}_{n}$ and $\ket{e}_{n}$ respectively. Here, and from now on, units of $\hbar=c=1$ have been taken. 

We are interested in the regime where all atoms are confined in a region much smaller than the wavelength $\lambda_0=2\pi/\omega_0$ of their transition, so that we can neglect any retardation effect and assume that all atoms see the same phase for the electromagnetic cavity field \cite{MaWo95,BrPe02,Gonzalez25}. In such a case, the interaction Hamiltonian under the rotating wave approximation in the interaction picture is given by  
\begin{equation}\label{hamiltonian}
		H_{I}(t)=\displaystyle\sum_{\bm{k}}g(\omega_{k})\left[J_{-}a^{\dagger}_{\bm{k}}\ee^{-\ii\Delta_{k}t}+\text{H.c.}\right],
\end{equation}
where $g(\omega_k)$ is a coupling function assumed to be real and dependent only on the frequency of the mode $\omega_k$, $J_{-}=\sum_{n=1}^{N}\sigma_{-}^{(n)}$ with ${\sigma_{-}^{(n)}=\ket{g}_{n\,n}\!\bra{e}}$, and ${\Delta_{k}=\omega_{k}-\omega_0}$. This form of the interaction is valid when the atomic transition frequency $\omega_0$ is much larger than the collective coupling strength and the relevant frequency scales of the reservoir, so that the counter-rotating terms oscillate rapidly and can be neglected. An important property of this Hamiltonian is that it preserves the total number of excitations $\sum_{\bm{k}}a^{\dagger}_{\bm{k}}a_{\bm{k}}+\hat{n}$, with $\hat{n}=\sum_{m=1}^{N}\ket{e}_{m\,m}\!\bra{e}$ the total number of excited two-level atoms.

We shall assume that the cavity is initially in the vacuum state and our focus will be on the behavior of the radiated intensity
\begin{equation}\label{schro}
		I(t)=-\omega_{0}\dfrac{d}{dt}\langle \hat{n}\rangle(t)=-\omega_0\dfrac{d}{dt}\langle J_{z}\rangle(t),
\end{equation}
in the continuum limit $\sum_{\bm{k}}\rightarrow \lim_{V\to\infty}V/(4\pi^3
)\int\,d^3k$ (with $V$ the quantization volume), with a spectral density $\mathcal{J}(\omega):=\frac{\omega^2}{\pi^2}\lim_{V\to\infty}g^2(\omega)V$ given by the usual Lorentzian function describing decay of the resonant cavity mode,
\begin{equation}\label{spectraldensity}
    \mathcal{J}(\omega)=\frac{1}{2\pi}\frac{\gamma_0^2 \lambda}{(\omega-\omega_0)^2+\lambda^2}.
\end{equation}
Here $\gamma_0$ and $\lambda$ account for the strength and the frequency width of the coupling with the bosonic reservoir, respectively. The limit $\lambda\rightarrow0$ corresponds to a perfect (lossless) cavity. This form of the spectral density corresponds to approximating the real, multipeaked spectral distribution of the cavity around a single dominant resonance near the atomic frequency $\omega _{0}$. This assumption is appropriate provided that $\lambda/\omega_0$ is sufficiently small, which ensures a narrow resonance that concentrates its contribution within the positive part of the spectrum and avoids overlapping with neighboring peaks.

If we consider the weak-coupling limit, we can formulate a Markovian master equation governing the evolution of the density matrix of the $N$ atoms in the interaction picture \cite{BrPe02,Libro},
\begin{equation}\label{masterequation}
\dfrac{d\rho_{S}(t)}{dt}=\gamma_{M}\left[J_{-}\rho_{S}(t)J_{+}-\dfrac{1}{2}\left\{J_{+}J_{-},\rho_{S}(t)\right\}\right],
\end{equation}
with the Markovian decay rate $\gamma_{M}=2\pi \mathcal{J}(\omega_0)=\gamma_0^2/\lambda$. This approximation is expected to be accurate when the bath correlation time is small compared to the relaxation time of the system, which, for the Lorentzian spectral density \eqref{spectraldensity} is equivalent to $\gamma_0^2/\lambda^2\ll1$, known as the bad cavity limit. If it is assumed further that the number of atoms $N$ is large and all atoms are initially excited, the mean-field (or semiclassical) approximation may be taken \cite{Bonifacio71a,Bonifacio71b,MaWo95,BrPe02}, which leads to the following expression for the radiated intensity,
\begin{equation}\label{Isr}
    I_{M}(t)=\dfrac{\omega_0\gamma_{M}N^2}{4}\left[\cosh\left(\dfrac{\gamma_{M} N}{2}(t-t_{0})\right)\right]^{-2},
\end{equation}
where $t_0\simeq\log(N+1)/(N\gamma_{M})$ is the time at which the maximum emission occurs. This expression possesses the characteristic features of superradiance: the radiated intensity reaches a maximum proportional to $N^2$ at a finite time $t_0$, in contrast to the exponential decay law that one would obtain if independent atoms were considered. However, we are interested in exploring the radiation process under more general conditions, namely, without invoking the Markovian and mean-field approximations.

\section{Exact analytical solutions for the dynamics of one and two atoms in a resonant QED cavity} 
\begin{figure}[t]
	\includegraphics[width=\columnwidth]{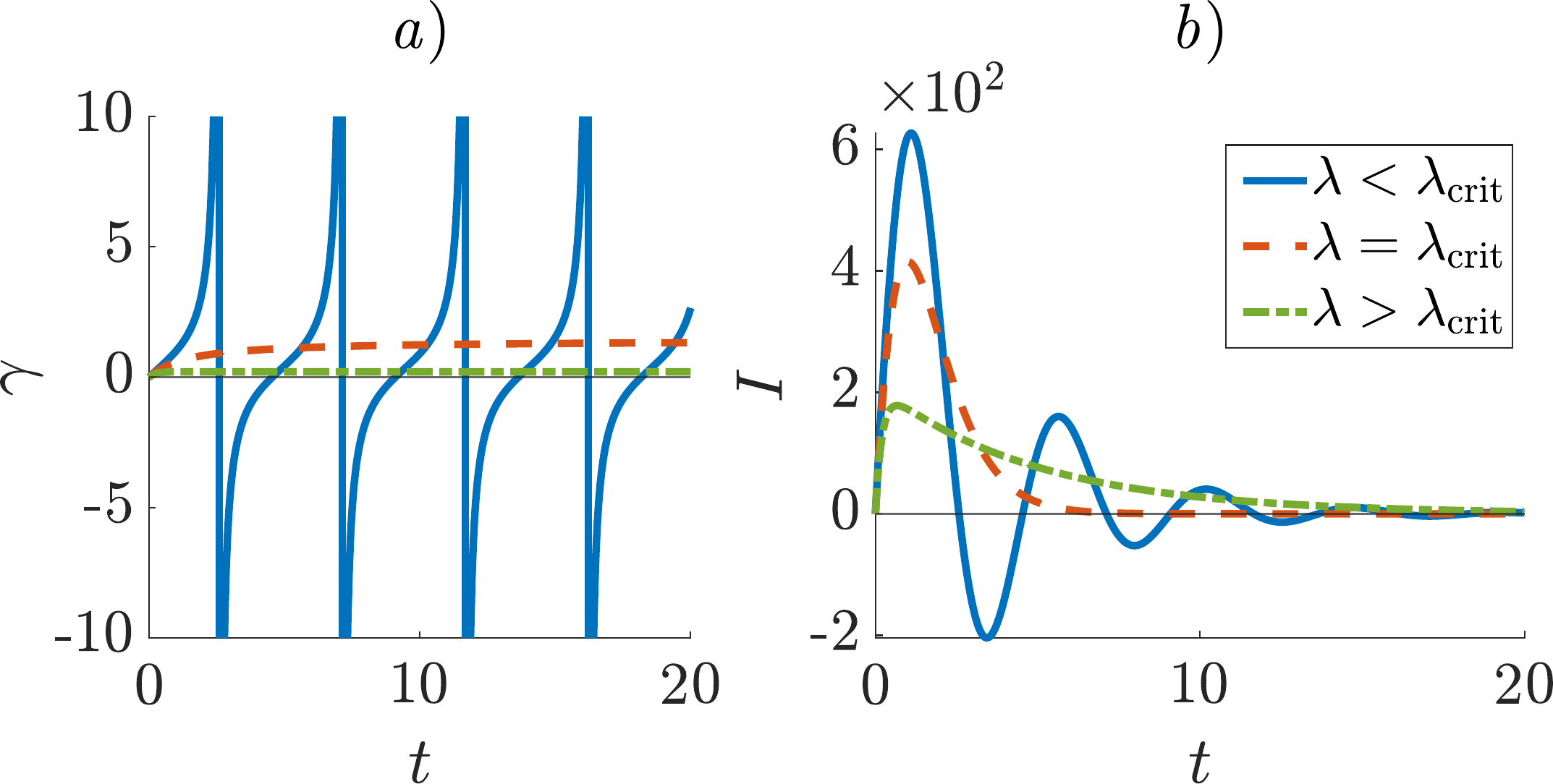}
	\caption{$\textit{a})$ Decay rate and $\textit{b})$ radiated intensity for a single atom, for $\lambda$ below $(\lambda=0.3)$, equal to, and above $(\lambda=5)$ the critical value $\lambda_{\rm crit}=\sqrt{2}$. The value $\omega_0=10^3$ was chosen. All quantities are in units of $\gamma_0=1$.}
	\label{Fig:1}
\end{figure}
The analytical exact solution of the previous model for $N=1$ is well-known \cite{BrPe02, Garraway97}, and it can be easily obtained as a solution of the non-Markovian master equation:
\begin{equation}
    \frac{d\rho(t)}{dt}\!=\!-\ii[H_S,\rho(t)]+\gamma(t)\!\big[\sigma_-\rho(t)\sigma_+-\tfrac12\{\sigma_+\sigma_-,\rho(t)\}\big].
\end{equation}
where the time-dependent decay rate is given by
\begin{equation}\label{dr1atom}
    \gamma(t)=\dfrac{2\gamma_0^2}{\lambda+\Omega_{1}\coth\left(\frac{\Omega_{1} t}{2}\right)},
\end{equation}
with $\Omega_{1}=\sqrt{\lambda^2-2\gamma_0^2}$. If ${\lambda<\sqrt{2}\gamma_0}$, $\gamma(t)$ becomes a periodic function with simple poles at $t_n=\frac{2}{|\Omega_1|} [\pi  n-\cot^{-1}(\tfrac{\lambda}{|\Omega_1|})]$, with $n=1,2,\ldots$ and takes negative values for $t\in(t_n,\frac{2\pi n}{|\Omega_1|} )$, see Fig. \ref{Fig:1}\textit{a}). The radiated intensity may be calculated analytically, yielding
\begin{multline}\label{emission1atom}
    I(t)=\dfrac{2\omega_0\gamma_0^2}{\Omega_1}\ee^{-\lambda t}\left[\cosh\left(\frac{\Omega_1t}{2}\right)\sinh\left(\frac{\Omega_1t}{2}\right)\right.\\+\left.\frac{\lambda}{\Omega_1}\sinh^2\left(\frac{\Omega_1t}{2}\right)\right].
\end{multline}
This function is shown in Fig.~\ref{Fig:1}\textit{b}). The maximum radiated emission occurs at time $t_{\rm max}=\frac{2}{\Omega_1}\tanh^{-1}\left(\frac{\Omega_1}{\sqrt{\lambda^2+2\gamma_0^2}}\right)$, with this maximum value given by
\begin{align}\label{maxemission1atom}
    I(&t_{\max})=\max_{t}I(t)\nonumber \\
    &=\frac{\omega_0}{2}\left(\lambda+\sqrt{\lambda^2+2\gamma_0^2}\right)\ee^{-\frac{2\lambda}{\Omega_1}\tanh^{-1}\left(\frac{\Omega_1}{\sqrt{\lambda^2+2\gamma_0^2}}\right)}.
\end{align}
For $\lambda>\sqrt{2}\gamma_0$, \eqref{emission1atom} is a positive function that increases monotonically from $I(0)=0$ up to the maximum value \eqref{maxemission1atom}, and from that point forward it decreases monotonically [green line in Fig.~\ref{Fig:1}\textit{b})]. On the other hand, if $\lambda<\sqrt{2}\gamma_0$, \eqref{emission1atom} becomes a damped periodic function oscillating between positive and negative values [blue line in Fig.~\ref{Fig:1}\textit{b})], with positive local maxima located at times $t_{\rm max}^{(n)}=\frac{2}{\lvert\Omega_1\rvert}\left[(n-1)\pi+\tan^{-1}\left(\frac{\lvert\Omega_1\rvert}{\sqrt{\lambda^2+2\gamma_0^2}}\right)\right]$, and negative local minima at $t_{\rm min}^{(n)}=\frac{2}{\lvert\Omega_1\rvert}\left[n\pi-\tan^{-1}\left(\frac{\lvert\Omega_1\rvert}{\sqrt{\lambda^2+2\gamma_0^2}}\right)\right]$. For those times at which $I(t)<0$, the atom is effectively reabsorbing the previously emitted radiation back from the field, with the maximum reabsorption occurring at $t_{\rm min}^{(1)}$, where
\begin{figure*}[t]
	\includegraphics[width=\textwidth]{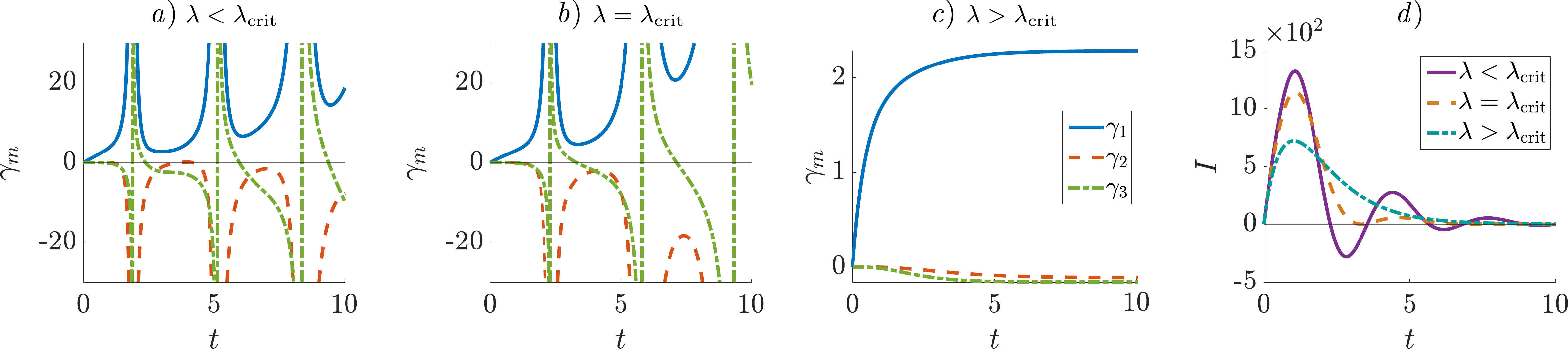}
	\caption{Canonical decay rates for $\textit{a})$ $\lambda<\lambda_{\rm crit}$ ($\lambda=1/2$), $\textit{b})$ $\lambda=\lambda_{\rm crit}$ ($\lambda=0.9024$) and $\textit{c})$ $\lambda>\lambda_{\rm crit}$ ($\lambda=2.3$), and corresponding radiated intensities $\textit{d})$. The value $\omega_0 = 10^3$ is chosen. At all times, at least one canonical decay rate is negative. For $\lambda<\lambda_{\rm crit}$ the radiated intensity also becomes negative during certain time intervals, indicating reabsorption. All quantities are in units of $\gamma_0=1$.}
	\label{Fig:2}
\end{figure*}
\begin{align}
    &I(t_{\rm min}^{(1)})=\min_{t}I(t)\nonumber\\
    &=\!-\frac{\omega_0}{2}\!\left(\sqrt{\lambda^2+2\gamma_0^2}-\lambda\right)\!\ee^{-\frac{2\lambda}{|\Omega_1|}\left[\pi-\tan^{-1}\left(\frac{|\Omega_1|}{\sqrt{\lambda^2+2\gamma^2}}\right)\right]}.
\end{align}
From these expressions we can see that as $\lambda$ approaches the critical value $\lambda_{\rm crit}=\sqrt{2}\gamma_0$ from below, $\min_{t}I(t)=I(t_{\rm min}^{(1)})\rightarrow0$ and $t_{\rm min}^{(1)}\rightarrow\infty$, i.e., the reabsorption decreases and is delayed [orange line in Fig.~\ref{Fig:1}\textit{b})]. Therefore, $\lambda_{\rm crit}$ separates the regimes with and without reabsorption.

To the best of our knowledge, a complete analytical solution has remained elusive for $N>1$, and only partial results are known by restricting the dynamics to the single-excitation subspace (see e.g. \cite{Gonzalez25,2atd0,Chruscinski23}). Here, we present a complete analytical solution for the case $N=2$, which will help to illustrate both the richness of the dynamics and the difficulty of the general problem.

Since the Hamiltonian \eqref{hamiltonian} is given in terms of collective angular momentum operators, it is convenient to introduce the Dicke state vectors:
\begin{align}\label{corr12}
\ket{N}&=\ket{e,e,\dots,e},\nonumber\\
\ket{N-1}&=\mathcal{S}\ket{g,e,\dots,e},\nonumber\\
\ket{N-2}&=\mathcal{S}\ket{g,g,e,\dots,e},\nonumber\\
&\hspace{5pt}\vdots \nonumber\\
\ket{1}&=\mathcal{S}\ket{g,\dots,g,e},\nonumber\\
\ket{0}&=\ket{g,g,\dots,g}\nonumber,
\end{align}
where $\mathcal{S}$ is the symmetrization operator, i.e. $\ket{m}$ represents the symmetric state of $m$ out of $N$ atoms in the excited state. The $J_{z}$ and $J_{\pm}$ operators act on these as
\begin{align}
		&J_{z}\ket{m} = \frac{2m-N}{2}\ket{m},\\
	    &J_{-}\ket{m} = \sqrt{m(N-m+1)}\ket{m-1},\\
        &J_{+}\ket{m}=\sqrt{(m+1)(N-m)}\ket{m+1}.
\end{align}
After a lengthy and technically involved analytical derivation, based on the solution of a coupled set of integro-differential equations and their careful treatment in the thermodynamic limit of the bath degrees of freedom, as detailed in Appendix~\ref{app:A}, we obtain the exact dynamics of the two atoms, and the master equation it satisfies in the interaction picture:
\begin{equation}\label{exactmasterequation}
    \dfrac{d\rho_S(t)}{dt}=\!\!\!\sum_{m,n=1}^{3}\!\!\!\Gamma_{mn}(t)\!\!\left[L_{m}\rho_S(t)L_{n}^{\dagger}\!-\!\tfrac{1}{2}\!\left\{L_{n}^{\dagger}L_{m},\rho_S(t)\right\}\right]\!,
\end{equation}
where the jump operators written in terms of Dicke states are
\begin{equation}
    \begin{cases}
        L_1=\ket{1}\bra{2},\\
        L_2=\ket{0}\bra{1},\\
        L_3=\ket{0}\bra{2}.
    \end{cases}
\end{equation}
The matrix $\Gamma_{mn}(t)$ is real, and its only nonzero entries are $\Gamma_{11}(t)$, $\Gamma_{12}(t)=\Gamma_{21}(t)$, $\Gamma_{22}(t)$, and $\Gamma_{33}(t)$. The specific expressions for these are found in the Appendix \ref{app:A} [see Eqs. \eqref{Gamma11}-\eqref{Gamma12}].

From this exact solution, we can calculate the radiation intensity when both atoms are initially excited, as shown in Fig.~\ref{Fig:2}\textit{d}). Similar to the case of a single atom, for small values of $\lambda$, the radiated intensity exhibits a damped oscillatory behavior [purple line in Fig.~\ref{Fig:2}\textit{d})]. Reabsorption of the emitted radiation in this regime constitutes a memory effect, and is a hallmark of non-Markovian dynamics \cite{BLP09,RHP10,revMarko1,revMarko2,revMarko3}. Correspondingly, we shall refer to an emission profile in which reabsorptions are present as non-Markovian. For large values of $\lambda$, the intensity displays a single delayed peak followed by a monotonic decay, resembling the typical superradiant profile. We shall refer to this behavior as Markovian; however, the form of the radiated emission profile alone is not sufficient to guarantee Markovianity without analyzing other dynamical properties. Again, as in the single-atom case, there exists a critical value of $\lambda$ that separates these two regimes, given by $\lambda_{\rm crit}=\sup\{\lambda|\min_t I(t)<0\}$. However, the behavior at this critical point differs markedly from that of a single atom: for $N = 2$, we observe pulsed emission, characterized by finite times at which $I(t) = 0$ [orange line in Fig.~\ref{Fig:2}\textit{d})]. At these times, the emission process halts and resumes again until it has decayed completely to the ground state. In this case, it is found that this critical value is approximately given by $\lambda_{\rm crit} \simeq 0.9024\gamma_0$. In order to determine this, a simple bisection (Bolzano-like) root-finding procedure was used to locate the value of $\lambda$ that makes zero the first local minimum of the radiated intensity since, in practice, due to the damped oscillatory nature of the intensity, examining the first local minimum is sufficient to determine the sign of $\min_tI(t)$.

The eigenvalues $\gamma_m(t)$ of $\Gamma_{mn}(t)$ in \eqref{exactmasterequation} are the canonical decay rates \cite{Michael14}, given in this case by
\begin{align}
        \gamma_{1,2}(t)&=\dfrac{1}{2}\bigg[\Gamma_{11}(t)+\Gamma_{22}(t)\nonumber\\
        &\qquad\qquad\pm\sqrt{\left[\Gamma_{11}(t)-\Gamma_{22}(t)\right]^2+4\Gamma_{12}^2(t)}\bigg],\\
        \gamma_{3}(t)&=\Gamma_{33}(t).
\end{align}
These are also plotted in Fig.~\ref{Fig:2}\textit{a})--c) for the corresponding values of $\lambda$. As we can see, in all cases at least one of the canonical decay rates is negative for any value of $t$ (we include an analytical proof of this for $\lambda\gg\gamma_0$ in Appendix \ref{app:B}). Since a negative canonical decay rate indicates non-Markovianity, the system possesses so-called eternal non-Markovianity \cite{Michael14,Megier17,Budini21}. This is an extreme form of non-Markovian evolution which has been recently observed in microscopic models for a qubit system using numerical simulations \cite{Amati24} at finite temperature, or perturbative approximation methods \cite{Gulacsi23}. We obtain it here in an exactly solvable model for a two-qubit system.

\begin{figure}[t]
	\includegraphics[width=\columnwidth]{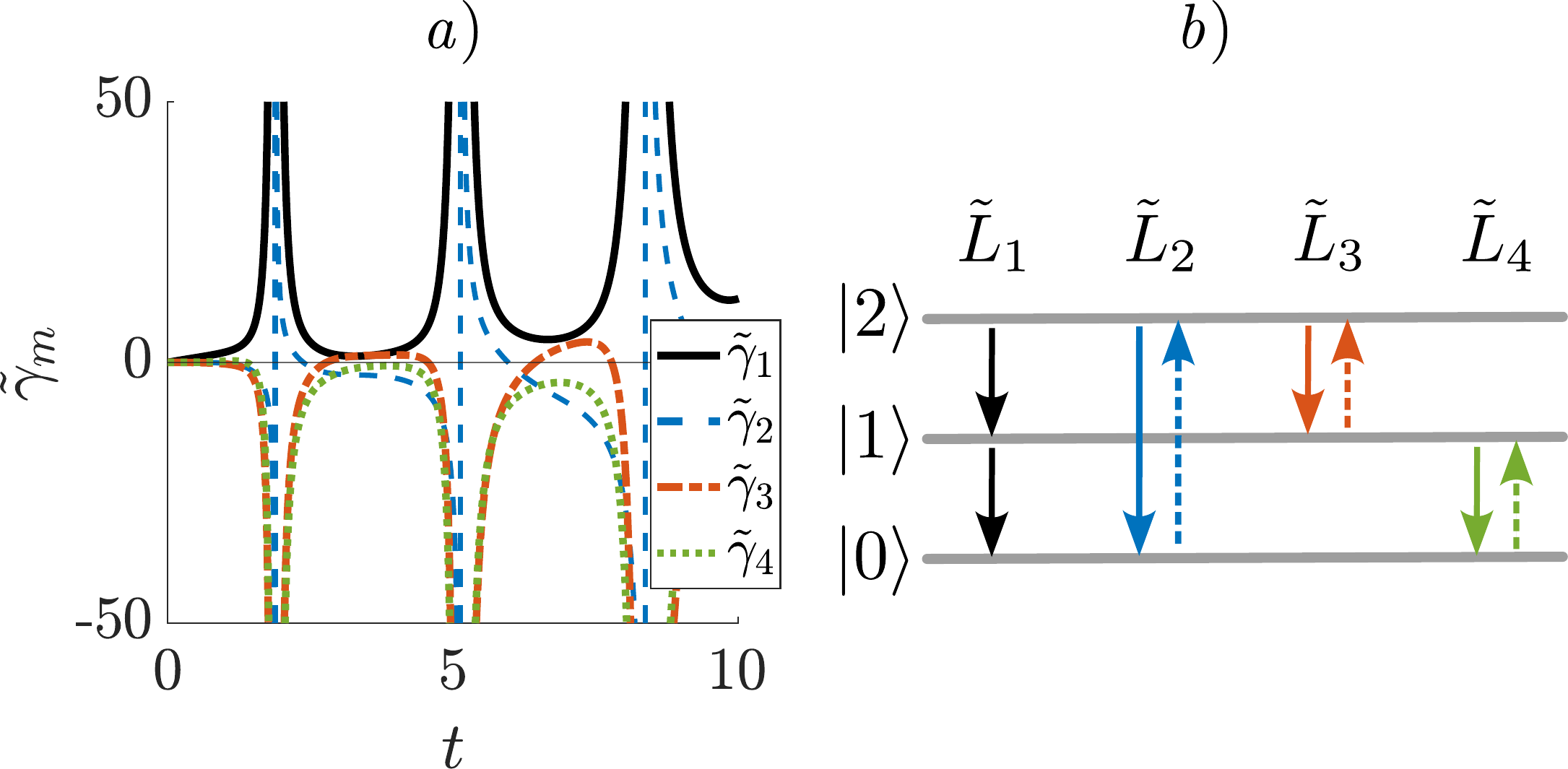}
	\caption{$\textit{a})$ Decay rates $\tilde{\gamma}_m(t)$ (non-canonical) for jump operators $\tilde{L}_m$ given by powers of $J_\pm$. We see that $\tilde{\gamma}_1$ remains positive for all $t$, leading to one-step decays throughout the Dicke state hierarchy ($\textit{b})$, black arrows). The remaining decay rates, associated with higher-order powers of $J_\pm$, can be positive or negative depending on time, effectively resulting in lowering ($\textit{b})$, solid arrows) and raising ($\textit{b})$, dashed arrows) processes across the Dicke state hierarchy, respectively. The same value of $\lambda$ as in Fig.~\ref{Fig:2}\textit{a}) has been taken. All quantities are in units of $\gamma_0=1$.}
	\label{Fig:3}
\end{figure}

In order to compare this exact equation with the Markovian approximation \eqref{masterequation}, we may alternatively rewrite \eqref{exactmasterequation} as
\begin{equation}\label{exactmasterequation2}
    \dfrac{d\rho_S(t)}{dt}=\!\!\sum_{m=1}^{4}\tilde{\gamma}_{m}(t)\!\left[\tilde{L}_{m}\rho_S(t)\tilde{L}_{m}^{\dagger}-\tfrac{1}{2}\big\{\tilde{L}_{m}^{\dagger}\tilde{L}_{m},\rho_S(t)\big\}\right]\!,
\end{equation}
with
\begin{align}
        \tilde{L}_1&=\ket{0}\bra{1}+\ket{1}\bra{2}=\tfrac{1}{\sqrt{2}}J_{-},\\
        \tilde{L}_2&=\ket{0}\bra{2}=\tfrac{1}{2}J_{-}J_{-},\\
        \tilde{L}_3&=\ket{1}\bra{2}=\tfrac{1}{2\sqrt{2}}J_{+}J_{-}J_{-},\\
        \tilde{L}_4&=\ket{0}\bra{1}=\tfrac{1}{2\sqrt{2}}J_{-}J_{-}J_{+},
\end{align}
and now
\begin{align}
        &\tilde{\gamma}_1(t)=\Gamma_{12}(t), &&\tilde{\gamma}_2(t)=\Gamma_{33}(t),\nonumber\\
        &\tilde{\gamma}_3(t)=\Gamma_{11}(t)-\Gamma_{12}(t), &&\tilde{\gamma}_4(t)=\Gamma_{22}(t)-\Gamma_{12}(t).
\end{align}
These (non-canonical) decay rates are plotted in Fig.~\ref{Fig:3}\textit{a}) for the same values $\lambda$ and $\gamma_0$ as in Fig.~\ref{Fig:2}\textit{a}). As can be seen, the decay rate $\tilde{\gamma}_1(t)$ associated with the jump operator $J_{-}$ is positive at all times. Therefore, this term in \eqref{exactmasterequation2} cannot be responsible for the reabsorption observed in the system, contrary to what one might expect from the $N=1$ case. Hence, in the non-Markovian regime, having $N>1$, it is not sufficient to describe the dynamics using a master equation of the Markovian form \eqref{masterequation} with the decay rate $\gamma_M$ simply promoted to a time-dependent one. To correctly capture the atomic dynamics in the non-Markovian regime, higher-order powers of $J_\pm$ must appear in the jump operators $\tilde{L}_i$ of the master equation.

\section{Spontaneous emission of \texorpdfstring{$N$}{N} atoms in a resonant QED cavity}
The exact master equation for $N=2$ atoms, strongly suggests that for an arbitrary number $N$ of atoms, the dynamics is solution to an equation with the same form as  \eqref{exactmasterequation} with $L_m$ the $N(N+1)/2$ jump operators between Dicke states, $\ketbra{i}{j}$, with $i<j$. However, the process of finding the coefficients $\Gamma_{mn}(t)$ in the thermodynamic limit of the bath degrees of freedom can be tedious as $N$ increases. Thus, in order to circumvent this problem and find the exact dynamics for an arbitrary number of atoms $N$ we shall adopt a different strategy.

In the regime where the Hamiltonian \eqref{hamiltonian} is valid, i.e. where all atoms are close enough to see the same phase for the electromagnetic cavity, there is only one bath correlation function given by $f(t-t')=\frac{\gamma_0^2}{2}\ee^{-\lambda|t-t'|}$ [see \eqref{corrfunction}]. This exponential form, independent of any distance between atoms, is the same as for the case of a single atom, and admits a mapping to a different environmental model where the reservoir is replaced by a single bosonic mode, called the pseudomode, with creation operator $b^{\dagger}$ and initially in the ground state, which interacts with the system via some Hamiltonian $H_{SP}$, and is damped by a time independent dissipator with the form $\mathcal{D}(\cdot)=b(\cdot)b^\dagger-\frac12\{b^\dagger b,\cdot\}$ \cite{Garraway96,Garraway97,2atd02,Tamascelli18,Lambdert19,Pleasance20}. While such a mode is generally a formal mathematical construction—hence the prefix ``pseudo''—in this specific setup it can be identified as a physical, isolated cavity mode subject to mirror leakage, as illustrated in Fig.~\ref{Fig:0}. More specifically, we obtain the equivalent master equation for the atoms-pseudomode system in the interaction picture,
\begin{align}\label{masterequationPM}
\dfrac{d\rho_{SP}(t)}{dt}=-&\ii\left[H_{SP},\rho_{SP}(t)\right]\nonumber \\
&+2\lambda\left[b\rho_{SP}(t)b^{\dagger}-\dfrac{1}{2}\left\{b^{\dagger}b,\rho_{SP}(t)\right\}\right],
\end{align}
where now

\begin{equation}\label{hamiltonian2}
		H_{SP}=\frac{\gamma_0}{\sqrt{2}}\left[J_{-}b^{\dagger}+J_{+}b\right]
\end{equation}
is the so-called Tavis-Cummings Hamiltonian \cite{TavisCummings68, Larson24}. In addition, we shall assume that all atoms are initially in their excited states $|\psi (0)\rangle=|e,e,\ldots,e\rangle=\ket{N}$. Since \eqref{masterequationPM} preserves the Dicke state subspace, the mapping reduces the initial infinitely-continuous dimensional unitary problem into a non-unitary one with density matrices of length $(N+1)^2$. This number may be reduced even further by noticing that $V=\ee^{\ii(\hat{n}+b^{\dagger}b)}$ is a weak symmetry of \eqref{masterequationPM}  \cite{Buca12, Gong16, Freter25}. Indeed, due to this symmetry, the subspace spanned by the set of eigenoperators of the superoperator $\mathcal{S}(\cdot)=[\hat{n}+b^{\dagger}b,\cdot]$ corresponding to the same eigenvalue is invariant under \eqref{masterequationPM}. For our initial condition $\mathcal{S}(\ket{N}\bra{N}\otimes\ket{0}\bra{0})=0$, so the only components of $\rho_{SP}(t)$ that may change under \eqref{masterequationPM} satisfy
\begin{equation}
    \mathcal{S}(\ket{i}\bra{j}\otimes\ket{l}\bra{m})=(i+l-j-m)\ket{i}\bra{j}\!\otimes\!\ket{l}\bra{m}=0.
\end{equation}
Hence, if we write $M=i+l$, and since $i$, $j$, $l$ and $m$ are non-negative, the only non-zero terms of $\rho_{SP}(t)$ belong to some ${W_{M}:={\rm{span}}\left\{\ket{i}\bra{j}\otimes\ket{l}\bra{m};i+l=j+m=M\right\}}$, which forms a $(M+1)\times(M+1)$ block. Additionally, we observe that for $A\in W_M$, $[H_{SP},A]\in W_{M}$ and $\mathcal{D}[A]\in W_M\oplus W_{M-1}$. Since the initial state is an element of $W_{N}$, it follows that $M\leq N$. Thus, at all times $t$ the solution of \eqref{masterequationPM} may be cast in block diagonal form ${\rho_{SP}(t)=\bigoplus_{M=0}^{N}\rho_{SP}^{(M)}(t)}$ with $\rho_{SP}^{(M)}(t)\in W_{M}$. The number of rows and columns of the density matrix is consequently reduced to $\sum_{M=0}^{N}(M+1)=(N+1)(N+2)/2$, and the number of non-zero elements is given by at most $\sum_{M=0}^{N}(M+1)^2=(N+1)(N+2)(2N+3)/6$.

This technique is benchmarked with the exact solution obtained for $N=2$ in appendix C, obtaining perfect agreement.

We proceed to solve the dynamics by numerically integrating \eqref{masterequationPM}. Let us note that the computational resources needed to do this grow cubically with the number of two-level atoms $N$. This will allow us to obtain exact results up to systems of $10^3$ atoms, which is an exceptionally large system for a non-Markovian exact simulation with the current state of the art \cite{Muller25}. For such system size, most of the studies are based on approximations, such as mean-field or higher-order cumulants truncations \cite{Fazio24}, e.g, setting
\begin{multline}\label{cummulants}
    \langle O_1O_2O_3\rangle=\langle O_1O_2\rangle\langle O_3\rangle+\langle O_1O_3\rangle\langle O_2\rangle\\+\langle O_1\rangle\langle O_2O_3\rangle-2\langle O_1\rangle\langle O_2\rangle\langle O_3\rangle,
\end{multline}
for any three observables $O_1$, $O_2$ and $O_3$ for the atoms and the pseudomode (and similarly at higher orders). This can be used to transform the linear master equation \eqref{masterequationPM} into a system of coupled nonlinear differential equations for the expectation values, the number of which does not increase with $N$. This procedure amounts to discarding higher order correlations among the different parts of the system. When the coupling of each individual atom to the mode is scaled as $N^{-1/2}$, these correlations decrease with higher $N$, so the method becomes exact in the limit $N\rightarrow\infty$. However, as we do not employ such a rescaling, the accuracy of this approximation for large $N$ is not guaranteed. For this reason, we proceed with the numerically exact method.

\subsection{Radiation regimes}

By fixing the value of $\gamma_0$, we explore different radiation regimes, evaluating numerically the radiated intensity $I(t)=-\omega_0\Tr\left[\hat{n}\dot{\rho}(t)\right]$ for different values of $\lambda$ and $N$.

\begin{figure}[t]
	\includegraphics[width=\columnwidth]{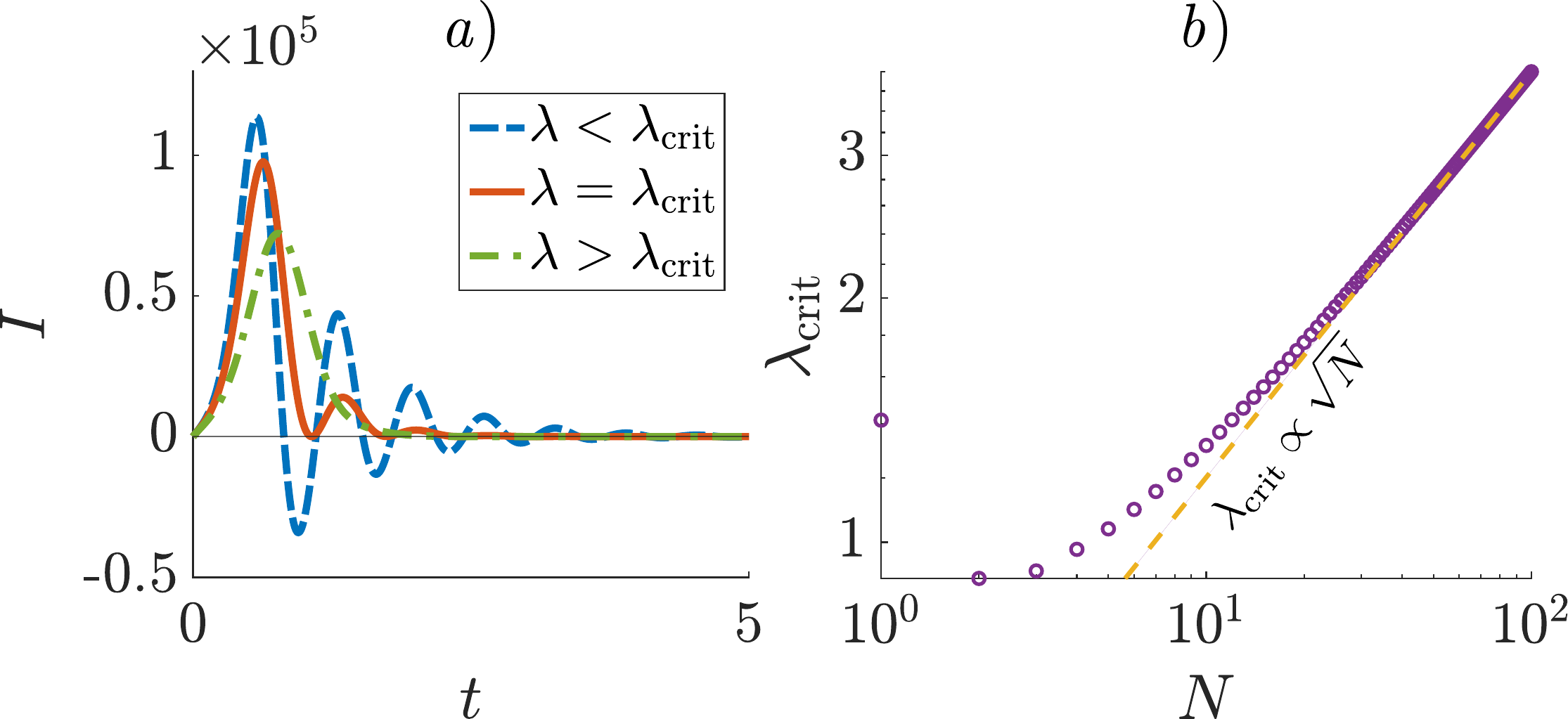}
	\caption{\textit{a}) Radiated intensity ($N=50$) for $\lambda$ below ($\lambda=1.3$), above ($\lambda=5.4$) and equal to the critical value $\lambda_{\rm crit}=2.7089$ (left), and \textit{b}) $\lambda_{\rm crit}$ as a function of $N$ (right). For $\lambda>\lambda_{\rm crit}$ (green line) a single burst is observed. For $\lambda<\lambda_{\rm crit}$ (blue line) there are times at which $I<0$, indicating reabsorption of excitations by the atoms. At $\lambda=\lambda_{\rm crit}$ (red line) pulsed emission without reabsorptions is observed. The value of $\lambda_{\rm crit}$ increases monotonically with $N$ from $N=2$ to $N=100$ (right plot) and appears to follow a $\sqrt{N}$ scaling law for large $N$. In these figures, the value $\omega_0 = 10^3$ is taken. All quantities are in units of $\gamma_0 = 1$.}
	\label{Fig:4}
\end{figure}

\begin{figure}[t]
	\includegraphics[width=\columnwidth]{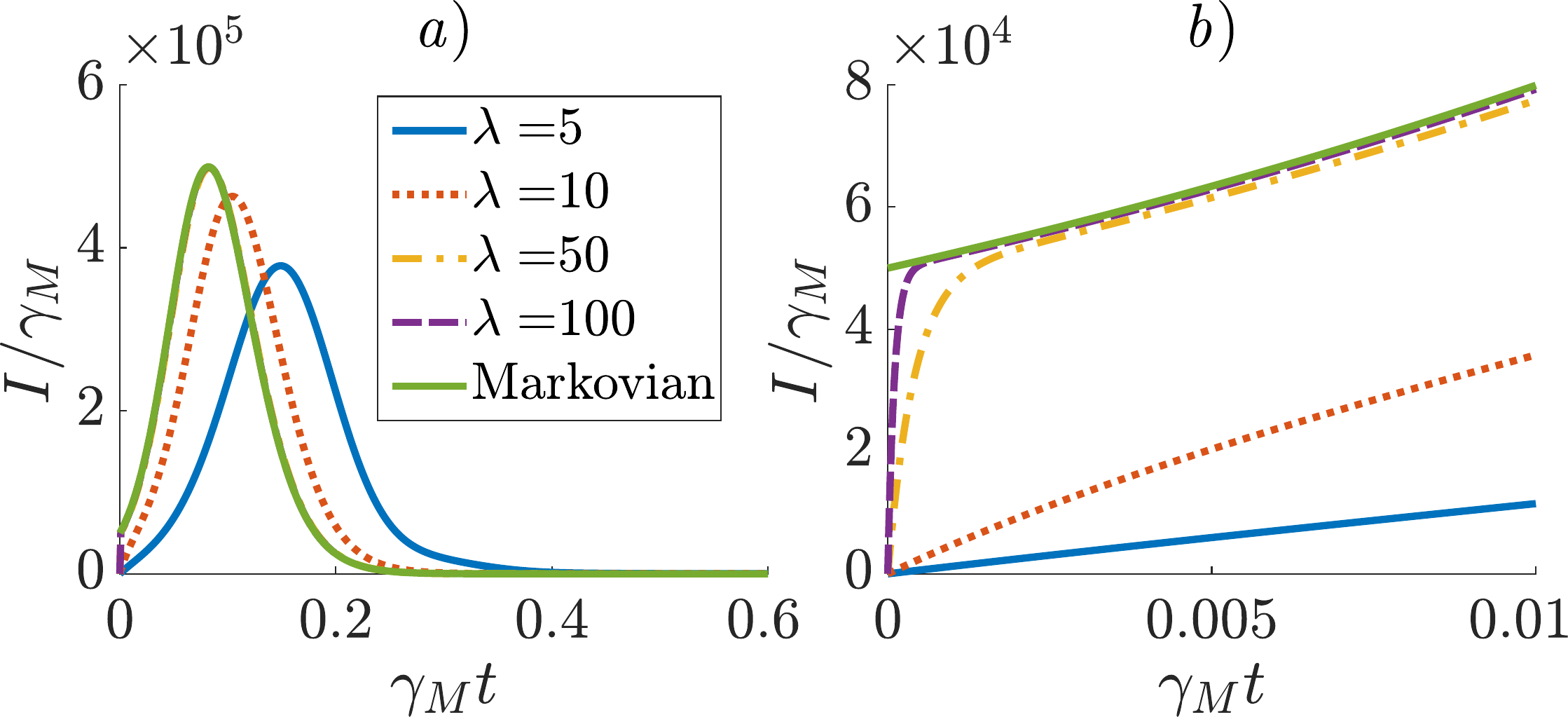}
	\caption{\textit{a}) Comparison of the radiated intensity ($N=50$) for increasing values of $\lambda$ with the Markovian result given by \eqref{masterequation} (green line), and \textit{b})  short time behavior of the same quantity. For these plots $\omega_0 = 10^3$. All quantities are in units of $\gamma_0 = 1$.}
	\label{Fig:8}
\end{figure}

For relatively large values of $\lambda$, we obtain the usual Markovian superradiant profile, with a maximum at a finite time, in contrast to the exponential decay law characteristic of independent atoms [see Fig.~\ref{Fig:4}\textit{a})]. As can be seen in Fig.~\ref{Fig:8}, as we increase $\lambda$ further, we find good agreement with the Markovian prediction given by \eqref{masterequation}. Note that as $\lambda$ increases, the radiated intensity becomes discontinuous at $t=0$, a characteristic feature of time-independent Markovian dynamics.

On the opposite end, for small values of $\lambda$ we observe a non-Markovian intensity profile. This is not surprising, considering that the extreme case $\lambda=0$ corresponds to the Tavis-Cummings model, which is unitary, as is evident from \eqref{masterequationPM}, and consequently has a quasi-periodic intensity profile.

As the intensity profile varies continuously with $\lambda$, one may expect the existence of a critical value $\lambda_{\rm crit}$ marking the crossover between the two types of behavior described, similar to the cases found in the previous section for two emitters. This is in fact what is obtained in Fig.~\ref{Fig:4}\textit{a}) (red line). At this critical value, pulsed emission is observed for all $N>1$, with emission peaks occurring at progressively shorter times. Using the same bisection procedure described for the two-atom case, we find $\lambda_{\rm crit}$ as a function of $N$ as shown in Fig.~\ref{Fig:4}\textit{b}). We see that  $\lambda_{\rm crit}$ starts at $\sqrt{2}\gamma_0$ for a single atom, drops for $N=2$, and increases monotonically from that point forward, at least as far as the numerical data shows. This implies that care should be taken when considering experimental setups, e.g., if one is trying to engineer Markovian emission by tuning the values of $\lambda$ and $\gamma_0$, paying no attention to the number of atoms of the cloud. Indeed, for fixed values of these parameters, the boundary between the Markovian and non-Markovian regimes depends on the number of atoms, so that increasing $N$ can place the system in the non-Markovian regime. This transition can be understood as follows: as explained later [see Eq.~\eqref{tau_R}], for a fixed $\gamma_0$ the effective collective coupling of the emitters to the environment scales as $\sqrt{N}\gamma_0$. Hence, even for fixed environmental parameters, the weak-coupling condition characteristic of the Markovian regime progressively deteriorates as the number of emitters increases. In fact, the available numerical results for $\lambda_{\rm crit}$ are consistent with $\lambda_{\rm crit}\propto\sqrt{N}$ for large $N$ [yellow-dashed line in Fig.~\ref{Fig:4}\textit{b})].

\subsection{The maximum of the radiated intensity} 

Figure~\ref{Fig:4}\textit{a}) suggests that, for a fixed number of atoms, the maximum of the radiated intensity increases as the memory effects grow. This is the behavior observed for a large number of $\lambda$ values, as shown in Fig.~\ref{Fig:5}\textit{a}). For a fixed $N$, the maximum emitted intensity is obtained in the Tavis–Cummings limit  ($\lambda \rightarrow 0$) and decreases monotonically as $\lambda$ increases.

This result may seem paradoxical considering the following: in the Markovian approximation, the maximum of the radiated intensity increases quadratically with the number of atoms $N$ as $N\rightarrow\infty$. We may ask if such an asymptotic power law holds in the opposite end, i.e., in the Tavis-Cummings limit. In order to answer this, let us consider a set of values for the number of atoms $\left\{N_{m}\right\}_{m=1}^{m_{\rm max}}$, and their corresponding maximum intensities, $\left\{\max \left[I(N_{m})\right]\right\}_{m=1}^{m_{\rm max}}$, given $\lambda$ and $\gamma_0$. We define the set $\left\{\nu_{m}\right\}_{m=1}^{m_{\rm max}-1}$ as 
\begin{equation}\label{exponent_def}
    \nu_{m}:= \dfrac{\log\left\{\max\left[I(N_{m+1})\right]/\max\left[I(N_{m})\right]\right\}}{\log(N_{m+1}/N_{m})}.
\end{equation}
This stands for a local estimation of the exponent of the maximum radiated intensity through a logarithmic slope. Indeed, it is clear that if an asymptotic law of the form $\max\left[I(N)\right]\propto N^{\mu}$ holds, then $\lim_{m\rightarrow\infty}\nu_{m}=\mu$. This is calculated for different values of $\lambda$ [see Fig.~\ref{Fig:5}\textit{b})]. As we can see, for large $\lambda$ we find $\nu\rightarrow2$, as expected. However, in the Tavis-Cummings limit we obtain $\nu\simeq1.5$. Consequently, since the maximum radiated intensity grows faster for larger values of $\lambda$, one should expect that the monotonically decreasing behavior of $\max(I)$ with $\lambda$ shown in Fig.~\ref{Fig:5}\textit{a}) will no longer hold when a sufficiently large $N$ is considered.

However, a more refined analysis raises questions about this conclusion. In the previous reasoning, there is the assumption that if the Markovian approximation holds for certain values of the environment parameters $\lambda$ and $\gamma_0$ and a small number of atoms $N$, then it will hold for any value of $N$. This is clearly not the case, since the transition between the Markovian and non-Markovian regimes may depend on $N$, as discussed above. 

To see analytically why this is the case, let us recall that the Markovian approximation is expected to hold when the bath correlation time—which in the present case is of order $\lambda^{-1}$—is much shorter than the characteristic relaxation time of the system. One might then be tempted to identify the latter with $\gamma_M^{-1}$, once the Markovian approximation is made in \eqref{masterequation}. However, to make a meaningful comparison across different values of $N$, one must also take into account the scaling of the jump operators. Since the characteristic relaxation timescales of the processes described by \eqref{masterequation} are set by the inverse of the real parts of $\gamma_M \lambda_n$, where $\{\lambda_n\}_{n=1}^{2^{2N}}$ are the eigenvalues of the generator $\mathcal{D}(\cdot)=J_{-}(\cdot)J_{+}-\frac{1}{2}\{J_{+}J_{-},\cdot\}$, all having negative real parts \cite{Libro}, an estimate of the average system relaxation time $\tau_R$ within the Markovian approximation is given by the inverse of the mean relaxation rate:
\begin{figure}[t]
	\includegraphics[width=\columnwidth]{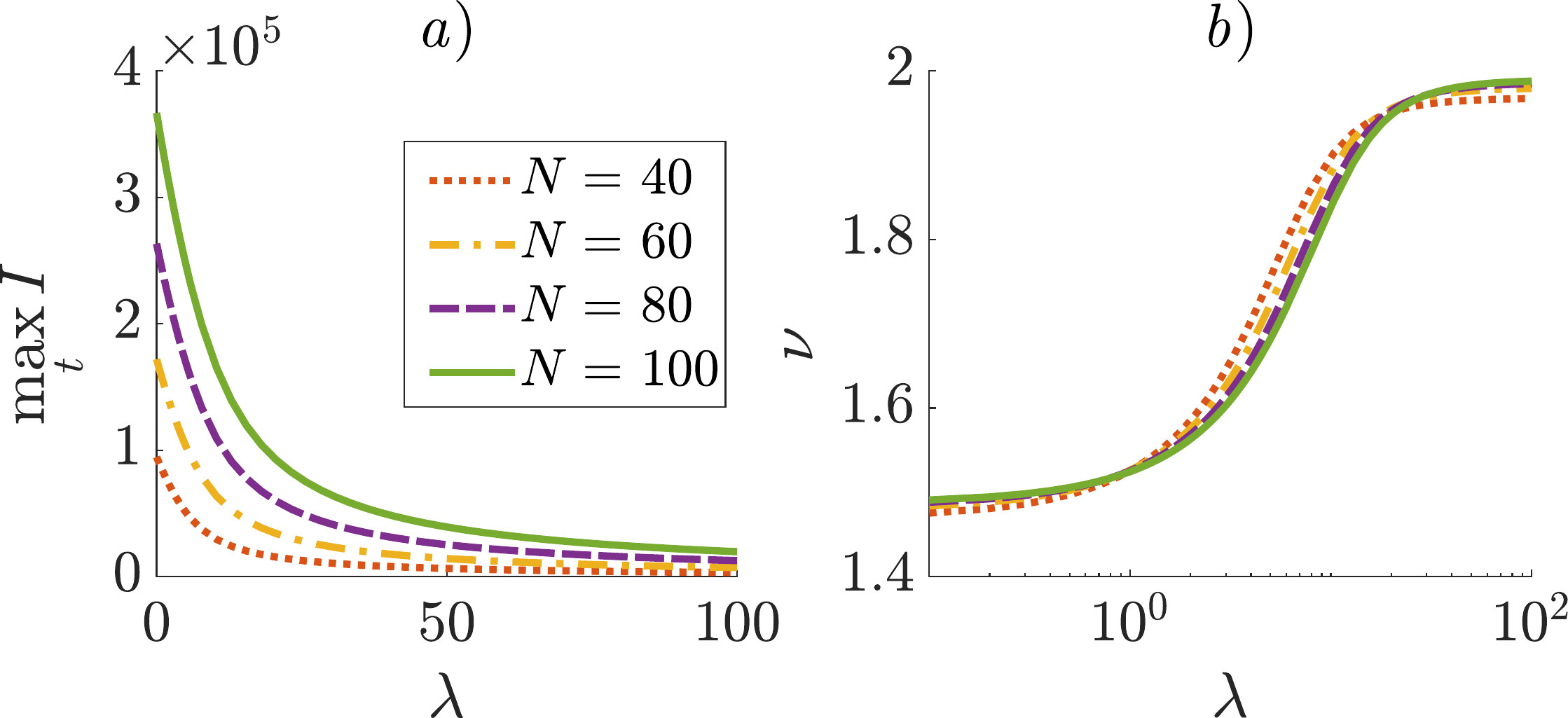}
	\caption{\textit{a}) Maximum of the radiated intensity and \textit{b}) local exponent \eqref{exponent_def} as functions of $\lambda$ for different values of $N$. The local exponent was calculated with $N_m$ in \eqref{exponent_def} given by the values in the legend, and $N_{m+1}$ given by 45, 65, 85 and 105, respectively. For these plots $\omega_0 = 10^3$. All quantities are in units of $\gamma_0 = 1$.}
	\label{Fig:5}
\end{figure}

\begin{align}\label{tau_R}
  \tau_R&=-\frac{1}{\gamma_M}\frac{2^{2N}}{\left(\sum_{n}\lambda_n\right)}=-\frac{2^{2N}}{\gamma_M \Tr(\mathcal{D})}\nonumber\\
  &=\frac{2^{2N}}{\gamma_M \{\Tr(J_+ J_-)-[\Tr(J_-)]^2\}}=\frac{2}{N\gamma_M}.
\end{align}
Having in mind that $\gamma_{M}=\gamma_0^2/\lambda$, this is equivalent to stating that the effective coupling strength of the collective system to the environment scales from $\gamma_0$ for a single emitter, to $\sqrt{N}\,\gamma_0$ for $N$. Consequently, the condition for the validity of the Markovian approximation becomes $\lambda \gg \sqrt{N}\,\gamma_0$. From this, it is clear that, for fixed environmental parameters, the approximation deteriorates as the number of two-level atoms increases. 

\begin{figure}[t]
	\includegraphics[width=\columnwidth]{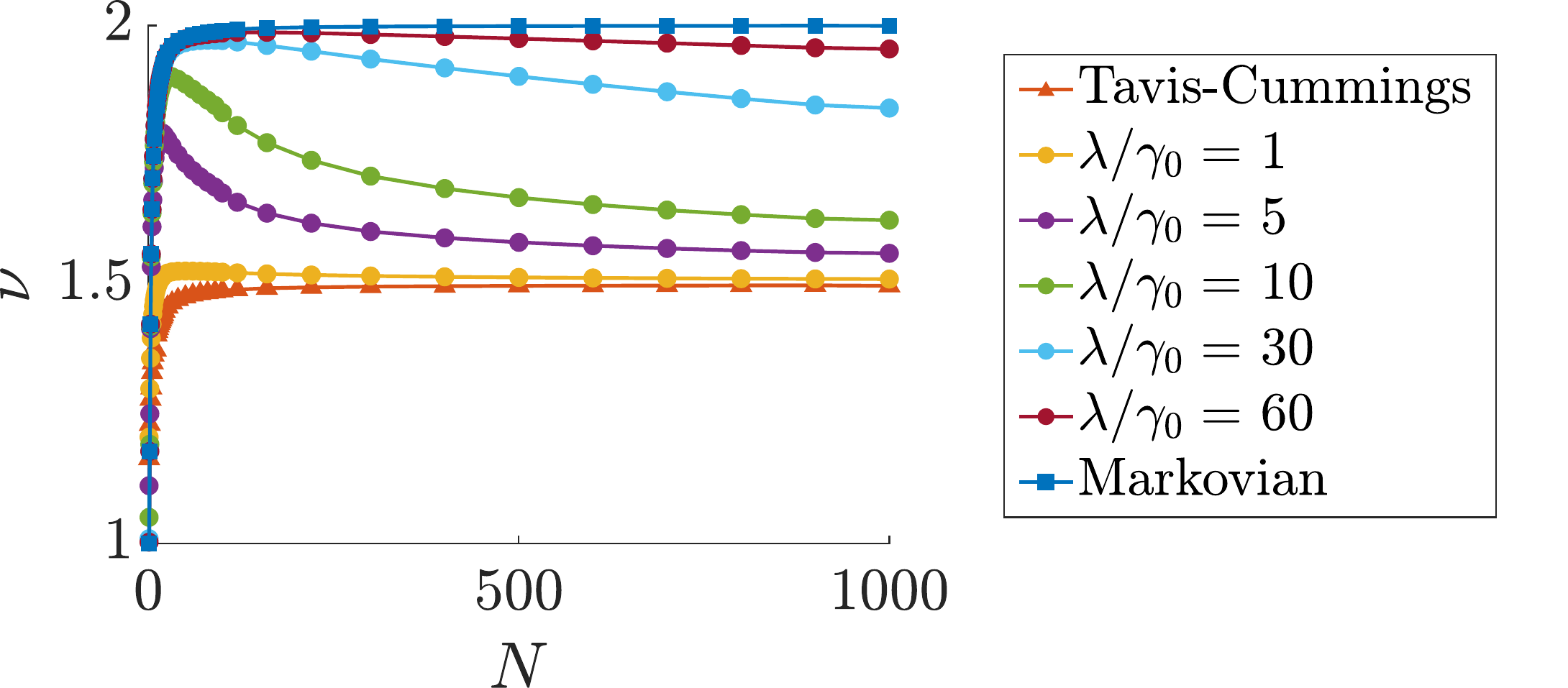}
	\caption{Local exponent \eqref{exponent_def} as a function of $N$ for different values of $\lambda/\gamma_0$, and $N$ up to $10^3$. Each data point is placed above its value of $N_m$ in \eqref{exponent_def}, with $N_{m+1}$ the location of the next data point. For $N_m=10^3$, the local exponent is calculated with $N_{m+1}=1001$. In all cases, for larger values of $\lambda$, the exponent increases close to the Markovian value of 2, but later decays toward the Tavis-Cummings limiting value of $1.5$. As $\lambda$ is decreased, this decay is observed at smaller values of $N$. In this plot $\omega_0 = 10^3$. All quantities are in units of $\gamma_0 = 1$.}
	\label{Fig:6}
\end{figure}

As a result, the quadratic exponent observed in Fig.~\ref{Fig:5}\textit{b}) for the largest values of $\lambda$ may reflect the validity of the Markovian approximation only at relatively small $N$. To investigate this further, in Fig.~\ref{Fig:6} we present numerically exact results for the local exponent defined in \eqref{exponent_def}, considering ensembles of up to $10^3$ two-level atoms. Indeed, we observe that for large $\lambda$, the exponent initially approaches the value $2$, but decreases as $N$ is further increased, eventually approaching the Tavis--Cummings limit of $1.5$.

This behavior can be understood as follows. Since the collective coupling of the atoms to the field increases with $N$, the characteristic timescale of the atomic dynamics decreases, so that photon emission becomes faster as the number of emitters grows. By contrast, the characteristic timescale over which the field dissipates these photons is set by the reservoir correlation time, $\lambda^{-1}$, and does not depend on $N$. As a consequence, for sufficiently large $N$, the atomic dynamics at short times becomes effectively faster than the dissipation process, and thus increasingly resembles the dissipation-free Tavis--Cummings model. Because the first maximum of the radiated intensity occurs within this short-time window, its scaling behavior approaches the Tavis-Cummings value in the large $N$ limit.

This is very remarkable as the usual quadratic superradiance scaling cannot be claimed as an asymptotic result, for large enough $N$. That scaling breaks down to $1.5$ unless the environmental correlation time or the system relaxation time are appropriately rescaled by a $N$ factor.

\subsection{The maximum of the reabsorbed intensity} 
\begin{figure}[t]
	\includegraphics[width=\columnwidth]{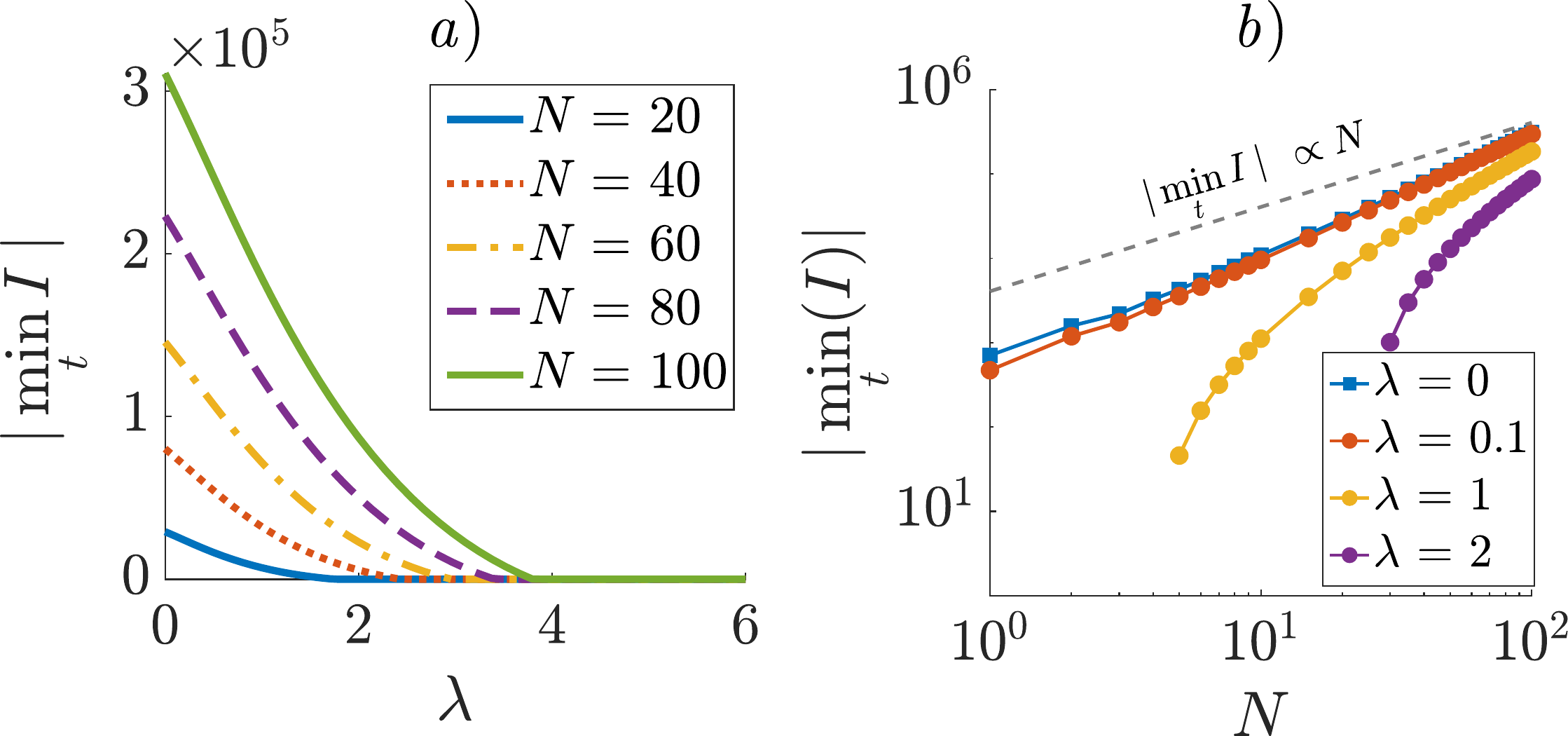}
	\caption{\textit{a}) Maximum reabsorbed intensity as a function of $\lambda$ for different values of $N$, and \textit{b}) as a function of $N$ for different values of $\lambda$. The maximum reabsorbed intensity decreases monotonically with $\lambda$ down to 0 at $\lambda = \lambda_{\rm crit}$, where it remains for larger values \textit{a}). As a function of $N$, we observe superlinear reabsorption in all cases, approaching the Tavis-Cummings exponent 1.5 for large $N$. In these figures $\omega_0 = 10^3$. All quantities are in units of $\gamma_0 = 1$.}
	\label{Fig:7}
\end{figure}

We may also consider the behavior of the maximum reabsorbed intensity $\lvert\min_tI(t)|$ as the values of $\lambda$ or $N$ are modified. As it is shown in Fig. \ref{Fig:7}\textit{a}), we see that for any value of $N$, maximum reabsorption decreases monotonically with $\lambda$. This is expected: as the value of $\lambda$ is reduced, the emitted photons remain longer inside the cavity, increasing the probability of reabsorption by the atoms. In addition, for a fixed value of $\lambda$, a larger number of two-level atoms produces more reabsorptions. 

As we did in the previous section for the maximum emission, we may ask if the maximum reabsorbed intensity follows a power law in the number of emitters. In order to study this, we plot in Fig. \ref{Fig:7}\textit{b}) the maximum reabsorbed intensity as a function of the number of atoms $N$ for different values of $\lambda$ using a log-log scale. If a power law holds for large $N$, the curves will approach straight lines, with the exponent given by their slopes. As can be seen, increasing $\lambda$ delays the convergence to a power-law behavior to larger values of $N$. For the largest values of $\lambda$, the local slopes of the curves decrease monotonically with $N$. Nevertheless, in all cases the curves tend toward the Tavis-Cummings result, which exhibits a slope of $1.5$ (the same scaling found for its maximum emission).

This behavior can be understood along lines similar to those discussed in the previous section for the decay of the emission exponent. For times satisfying $\lambda t \ll 1$, the dynamics is well approximated by the Tavis-Cummings model. Since the atomic dynamics accelerates as $N$ increases, for sufficiently large $N$ the first reabsorption peak occurs within this short-time regime. Consequently, its scaling behavior is well captured by the corresponding Tavis-Cummings prediction.

In order to compare this scaling behavior with the case of independent atoms, a segmented line for $\lvert\min_t I\rvert\propto N$ is included in Fig. \ref{Fig:7}\textit{b}). As we can see, in all cases the slopes are greater than $1$, indicating that the reabsorption displays collective features in this model. This is similar to the phenomenon of superabsorption \cite{Quach2022,Kamimura22,Higgins14,Yang21}, but with the key distinction that no external control or driving is required here. To distinguish this from externally engineered superabsorption, we refer to it as ``spontaneous superabsorption''.

\section{Conclusions} We have analyzed the exact spontaneous emission of an ensemble of $N$ identical and non-interacting two-level atoms coupled to a Lorentzian cavity. We carried out an analytical treatment of the $N=2$ case, deriving the full dynamics and the corresponding exact non-Markovian master equation. This analytical solution is especially significant given the well-known difficulty of obtaining closed-form results for non-Markovian dynamics, a challenge that becomes even more severe beyond the single-emitter regime. The exact solution reveals features absent in the single-emitter scenario, such as pulsed emission and a regime of eternal non-Markovianity, characterized by the persistent negativity of at least one canonical decay rate.

For larger ensembles, the exact analytical solution for $N=2$ already reveals the general structure of the non-Markovian master equation for arbitrary $N$. However, beyond the two-emitter case, the explicit analytical evaluation of the decay rates becomes prohibitively involved. To overcome this limitation, we show that the pseudomode method can be applied, and that the resulting master equation exhibits a weak symmetry that allows us to access the collective emission of large ensembles in a numerically exact manner, without resorting to mean-field or other approximation methods. Using this approach, we explored systems containing up to a $10^3$ emitters, an exceptionally large size for an exact non-Markovian treatment. Our analysis reveals three qualitatively distinct regimes of collective radiation governed by the spectral width $\lambda$: a Markovian regime with a single superradiant burst, a critical regime displaying periodic pulsed emission, and a non-Markovian regime dominated by coherent reabsorption of the emitted field.

The study of the maximum emitted intensity for the exact dynamics further reveals that the familiar superradiant scaling $I_{\rm max}\propto N^2$ progressively degrades with increasing $N$. As the collective coupling to the cavity field strengthens with the ensemble size, the effective system–bath interaction enters a regime in which emission becomes increasingly reabsorptive and the Markovian approximation breaks down. Consequently, the emission peak may, in some cases, transition smoothly from the quadratic Dicke scaling toward a subquadratic law that approaches the Tavis–Cummings value in the large $N$ limit. This behavior demonstrates that the cooperative enhancement of radiation is self-limiting: as the system becomes more collective, it also becomes more non-Markovian, and the emitted energy is partially retrieved rather than radiated away.

A particularly remarkable manifestation of this interplay is the occurrence of spontaneous superabsorption. In the non-Markovian regime, the atomic ensemble periodically reabsorbs part of the emitted field, leading to sharp revivals in the intensity, whose maximum scales superlinearly with $N$. This phenomenon originates from the coherent build-up of atomic excitations mediated by the cavity memory, effectively generating a collective feedback on the dynamics of the atoms.

In summary, our results provide an exact characterization of collective spontaneous emission bridging the Markovian and strongly non-Markovian limits. They show that environmental memory and atomic cooperativity are deeply intertwined, giving rise to qualitatively new regimes of radiation such as critical pulsed emission and spontaneous superabsorption. Beyond their fundamental significance, these phenomena suggest practical routes for exploiting non-Markovian cooperativity in engineered quantum systems, enabling controllable energy recycling, delayed superradiant pulses, or cavity-based superabsorptive devices.

\section*{Author Contributions} I.G. carried out the main analytical calculations, performed the numerical simulations, and prepared the figures. A.R. conceived the theoretical framework and scope of the work, contributed to the analytical developments, and supervised the project. Both authors discussed the results and participated in writing the manuscript.

A large language model tool was used exclusively for proofreading and improving the clarity of some parts of the manuscript. The authors take full responsibility for the work.

\begin{acknowledgments}
The authors acknowledge support from Spanish MICIN grants PID2021-122547NB-I00 and EUR2024-153545, from the TEC-2024/COM-84 QUITEMAD-CM project funded by Comunidad de Madrid, and from the ``MADQuantum-CM'' project funded by Comunidad de Madrid and by the Recovery, Transformation and Resilience Plan – Funded by the European Union - NextGenerationEU. I.G. acknowledges support from the MICIN contract PRE2022-101824 (MICIN/AEI/FSE,UE).
\end{acknowledgments}

\setcounter{secnumdepth}{2}
\setcounter{equation}{0}

\appendix

\renewcommand{\theequation}{\Alph{section}.\arabic{equation}}
\setcounter{equation}{0}

\bigskip

\section{Exact analytical solution for \texorpdfstring{$N=2$}{N=2}}\label{app:A}

We are assuming that the field is initially in the vacuum, so we want to solve the dynamics under the initial state condition
\begin{equation}
    \ket{\Psi(0)}=\ket{\psi(0)}\ket{0},
\end{equation}
with $\ket{\psi(0)}$ an arbitrary initial state of the atoms. The case of a mixed atomic initial state will follow from this solution by linearity.

\subsection{Dynamics of the system-environment composite}
For the case $N=2$ and since the total number of excitations is preserved by \eqref{hamiltonian}, the most general atoms-field state at an arbitrary time can be written in terms of Dicke states $\{\ket{0},\ket{1},\ket{2}\}$ in the form 
\begin{align}\label{state2atoms}
    \ket{\Psi(t)}=&c_{0}\ket{0}\ket{0}+c_{10}(t)\ket{1}\ket{0}+c_{-,0}\ket{-}\ket{0}\nonumber \\
    &+\sum_{\bm{k}}c_{0\bm{k}}(t)\ket{0}\ket{1_{\bm{k}}}+c_{20}(t)\ket{2}\ket{0}\nonumber \\
    &+\sum_{\bm{k}}c_{1\bm{k}}(t)\ket{1}\ket{1_{\bm{k}}}+\sum_{\bm{k}}c_{-,\bm{k}}\ket{-}\ket{1_{\bm{k}}}\nonumber\\
    &+\sum_{\bm{k}}c_{02\bm{k}}(t)\ket{0}\ket{2_{\bm{k}}}\nonumber\\
    &+\dfrac{1}{2}\sum_{\substack{\bm{k},\bm{k}' \\ \bm{k}\neq \bm{k}'}}c_{0\bm{k}\bm{k}'}(t)\ket{0}\ket{1_{\bm{k}}1_{\bm{k}'}},
\end{align}
where $\ket{-}:=\frac{1}{\sqrt{2}}(\ket{e,g}-\ket{g,e})$, $\ket{n_{\bm{k}}}$ denotes $n$ photons in the mode $\bm{k}$, and the $1/2$ factor in the last term is introduced for convenience, since the equivalent states $\ket{1_{\bm{k}}1_{\bm{k}'}}$ and $\ket{1_{\bm{k}'}1_{\bm{k}}}$ [and thus $c_{0\bm{kk}'}(t)=c_{0\bm{k}'\bm{k}}(t)$] appear as different terms in the summation. We now impose that $\ket{\Psi(t)}$ is a solution of the Schr\"odinger equation for the Hamiltonian $H_I(t)$ given in \eqref{hamiltonian}, $\frac{d}{dt}|\Psi(t)\rangle=-\ii H_I(t)|\Psi(t)\rangle$. Since the states $\ket{0}\ket{0}$, $\ket{-}\ket{0}$, and $\ket{-}\ket{1_{\bm{k}}}$ are invariant under this Hamiltonian, their corresponding coefficients in \eqref{state2atoms} do not exhibit any time dependence. The dynamics for the single initial excitation subspace is then given by components $\{\ket{1}\ket{0}$ and $\ket{0}\ket{1_{\bm{k}}}\}$ and can be easily solved. First, a formal integration of the $\ket{0}\ket{1_{\bm{k}}}$ coefficients in the Schr\"odinger equation yields
\begin{equation}\label{c0k}
    c_{0\bm{k}}(t)=-\ii\sqrt{2}g(\omega_{k})\int_{0}^{t}dt'\ee^{\ii\Delta_kt'}c_{10}(t').
\end{equation}
Substituting this expression into the differential equation for the coefficient of $\ket{1}\ket{0}$ leads to
\begin{equation}\label{integrodiff-solvable_0}
    \dot{c}_{10}(t)=-2\int_0^t dt' f(t-t')c_{10}(t'),
\end{equation}
where $f(t-t')$ is the bath correlation function
\begin{equation}\label{corrfunction}
    f(t-t'):=\sum_{\bm{k}} g^2(\omega_{k})\ee^{-\ii\Delta_{k}(t-t')}.
\end{equation}
In the thermodynamic limit (TL), $V\to\infty$, the modes form a continuum such that $\omega_k\rightarrow \omega$, and we have
\begin{equation}
    \sum_{\bm{k}}g^2(\omega_k)\xrightarrow{\rm TL}\dfrac{1}{4\pi}\int d^3k\frac{\mathcal{J}(\omega)}{\omega^2},
\end{equation}
where $\mathcal{J}(\omega):=\frac{\omega^2}{\pi^2}\lim_{V\to\infty}g^2(\omega)V$ is the spectral density. Since we are considering a Lorentzian spectral density \eqref{spectraldensity}, we may easily obtain the thermodynamic limit of the bath correlation function \eqref{corrfunction},
\begin{align}\label{corrfunctionth}
    f(t-t')&=\sum_{\bm{k}} g^2(\omega_{k})\ee^{-\ii\Delta_{k}(t-t')}\nonumber \\
    &\qquad\qquad\quad\xrightarrow{\rm TL} \dfrac{1}{4\pi}\int d^3k\frac{\mathcal{J}(\omega)}{\omega^2}\ee^{-\ii\Delta(t-t')}\nonumber \\
    &\simeq \int_{-\infty}^{\infty} d\omega \mathcal{J}(\omega)\ee^{-\ii\Delta(t-t')}=\frac{\gamma_0^2}{2}\ee^{-\lambda|t-t'|},
\end{align}
where $\Delta = \omega - \omega_0$. In the last step, we have assumed, as usual, that the Lorentzian is sufficiently narrow and centered around a frequency $\omega_0$ large enough to justify extending the integral over positive frequencies to the entire real line. After \eqref{corrfunction}, the equation \eqref{integrodiff-solvable_0} becomes
\begin{equation}\label{integrodiff-solvable}
    \dot{c}_{10}(t)=-\gamma_0^2\int_0^t dt' \ee^{-\lambda (t-t')}c_{10}(t').
\end{equation}
This integro-differential equation can be readily solved by taking time derivatives (or by using a Laplace transform), which allows it to be recast as the differential equation
\begin{equation}
    \ddot{c}_{10}(t)+\lambda \dot{c}_{10}(t) +\gamma_0^2 c_{10}(t)=0
\end{equation}
subject to the initial condition $\dot{c}_{10}(0)=0$. The solution is straightforward:
\begin{equation}\label{c10}
    c_{10}(t)=c_{10}(0)\upsilon(t),
\end{equation}
where 
\begin{equation}\label{sol1exc}
    \upsilon(t)=\ee^{-\frac{\lambda  t}{2}} \left[\cosh \left(\dfrac{\Omega t}{2}\right)+\dfrac{\lambda }{\Omega} \sinh \left(\dfrac{\Omega t}{2}\right)\right],
\end{equation}
and $\Omega=\sqrt{\lambda^2-4\gamma_0^2}$. This is completely analogous to the case of a single atom \cite{Garraway97,BrPe02}.

Therefore, what remains is to solve the dynamics within the subspace of two initial excitations, which is where the main difficulty lies. In this case, the Schr\"odinger equation leads to the following set of equations
\begin{align}
    &\ii\dot{c}_{20}(t)=\sqrt{2}\sum_{\bm{k}}g(\omega_{k})\ee^{-\ii\Delta_{k}t}c_{1\bm{k}}(t),\\
    \displaystyle &\ii\dot{c}_{1\bm{k}}(t)=\sqrt{2}g(\omega_{k})\ee^{\ii\Delta_{k}t}c_{20}(t)+2 g(\omega_{k})\ee^{-\ii\Delta_{k}t}c_{02\bm{k}}(t)\nonumber \\
    &\displaystyle\qquad\qquad\quad\quad\quad+\sqrt{2}\sum\limits_{\substack{\bm{k}' \neq \bm{k}}}g(\omega_{k'})\ee^{-\ii\Delta_{k'}t}c_{0\bm{k}\bm{k}'}(t),\\
    &\ii\dot{c}_{02\bm{k}}(t)=2\,g(\omega_{k})\ee^{\ii\Delta_{k}t}c_{1\bm{k}}(t),\\
    &\ii\dot{c}_{0\bm{k}\bm{k}'}(t)=\sqrt{2}\,g(\omega_{k'})\ee^{\ii\Delta_{k'}t}c_{1\bm{k}}(t)\nonumber \\
    &\qquad\qquad\qquad\qquad\quad\quad+\sqrt{2}\,g(\omega_{k})\ee^{\ii\Delta_{k}t}c_{1\bm{k}'}(t).
\end{align}
Since we take the field to be initially in the vacuum, we may integrate the last two equations formally as
\begin{align}	
    &c_{02\bm{k}}(t)=-\ii\,2g(\omega_{k})\int_{0}^{t}dt'\ee^{\ii\Delta_kt'}c_{1\bm{k}}(t'),\\
    &c_{0\bm{k}\bm{k}'}(t)=-\ii\sqrt{2}g(\omega_{k'})\int_{0}^{t}dt'\ee^{\ii\Delta_{k'}t'}c_{1\bm{k}}(t')\nonumber \\
    &\qquad\qquad\qquad\quad-\ii\,\sqrt{2}g(\omega_{k})\int_{0}^{t}dt'\ee^{\ii\Delta_kt'}c_{1\bm{k}'}(t').
\end{align}	
Inserting these in the remaining equations we obtain, after some algebra,
\begin{align}
    &\ii\dot{c}_{20}(t)=\sqrt{2}\sum_{\bm{k}}g(\omega_{k})\text{e}^{-i\Delta_{k}t}c_{1\bm{k}}(t),\label{finalsystem1} \\
    &\ii\dot{c}_{1\bm{k}}(t)=\sqrt{2}g(\omega_{k})\ee^{\ii\Delta_{k}t}c_{20}(t)+\ii w_k(t) \nonumber \\
    &\qquad\qquad\qquad\qquad\quad-4\ii\int_{0}^{t}dt'f(t-t')c_{1\bm{k}}(t'),\label{finalsystem2}
\end{align}
where $f(t-t')$ is the bath correlation function \eqref{corrfunctionth} and we have defined
\begin{multline}\label{aux_w}
    w_{k}(t)=2\int_{0}^{t}dt'f(t-t')c_{1\bm{k}}(t')\\
    -2\int_{0}^{t}dt'g(\omega_{k})\ee^{\ii\Delta_{k}t'}\Xi(t,t'),
\end{multline}
with
\begin{equation}\label{Pi}
    \Xi(t,t'):=\sum_{\bm{k}}g(\omega_k)\ee^{-\ii\Delta_kt}c_{1\bm{k}}(t').
\end{equation}
The usefulness of this apparently odd disposition of the terms will become clear in the following. We integrate \eqref{finalsystem2} by means of the Laplace transform $\mathfrak{L}$, obtaining
\begin{multline}
    \ii c_{1\bm{k}}(t)=\sqrt{2}g(\omega_k)\int_{0}^{t}dt'\phi(t-t')\ee^{\ii\Delta_{k}t'}c_{20}(t')\\
    +\ii\int_{0}^{t}dt'\phi(t-t')w_k(t').
\end{multline}
Here, $\phi(t)=\mathfrak{L}^{-1}\left[\frac{1}{s+4F(s)}\right](t)$, $\mathfrak{L}^{-1}$ is the inverse Laplace transform and $F(s):=\mathfrak{L}[f(t)](s)$. Inserting this into \eqref{finalsystem1}, we get
\begin{multline}\label{c110step2}
    \dot{c}_{20}(t)=-2\int_{0}^{t}dt'\phi(t-t')f(t-t')c_{20}(t')\\
    -\ii\sqrt{2}\int_{0}^{t}dt'\phi(t-t')\sum_{\bm{k}}g(\omega_{k})\ee^{-\ii\Delta_{k}t}w_k(t').
\end{multline}
We must now take the thermodynamic limit $V \to \infty$, which requires some care. From \eqref{corrfunctionth}, it follows immediately that
\begin{equation}
    F(s)=\mathfrak{L}\left[f(t)\right](s)\xrightarrow{\rm TL}\frac{\gamma_0^2}{2}\frac{1}{s+\lambda},
\end{equation}
in this limit. Consequently,
\begin{align}\label{phi(t)}
    \phi(t)&=\mathfrak{L}^{-1}\left[\frac{1}{s+4F(s)}\right](t)\nonumber\\
    &\quad\xrightarrow{\rm TL}\mathfrak{L}^{-1}\left[\frac{s+\lambda}{s(s+\lambda)+2\gamma_0^2}\right](t)\nonumber \\
    &\quad=\ee^{-\frac{\lambda t}{2}}\left[\cosh\left(\dfrac{\tilde{\Omega}t}{2}\right)+\dfrac{\lambda}{\tilde{\Omega}}\sinh\left(\dfrac{\tilde{\Omega}t}{2}\right)\right],
\end{align}
with $\tilde{\Omega}=\sqrt{\lambda^2-8\gamma_0^2}$. 

To calculate the thermodynamic limit of $\Xi(t,t')$, we begin by multiplying both sides of \eqref{finalsystem2} by $g(\omega_k)\ee^{-\ii \Delta_k \tau}$, with $\tau \geq t$, and summing over $\bm{k}$. Making use of Eqs.~\eqref{aux_w} and \eqref{Pi}, this leads to
\begin{multline}
    \!\!\!\!\partial_t{\Xi}(\tau,t)=-\ii\sqrt{2}f(\tau-t)c_{20}(t)-2\!\int_{0}^{t}\!\!dt'f(t-t')\Xi(\tau,t')\\-2\!\int_{0}^{t}\!dt'f(\tau-t')\Xi(t,t').
\end{multline}
We now take the thermodynamic limit $\Xi(t,t')\xrightarrow{\rm TL} \Xi_{\rm {TL}}(t,t')$. Assuming that this limit can be interchanged with time derivatives and integrals, and making use of \eqref{corrfunctionth}, the above equation becomes
\begin{multline}
		\partial_t{\Xi}_{\rm TL}(\tau,t)=-\ii\dfrac{\gamma_0^2}{\sqrt{2}}\ee^{-\lambda(\tau-t)}c_{20}(t)\\
        -\gamma_0^2\int_{0}^{t}dt'\ee^{-\lambda(t-t')}\Xi_{\rm TL}(\tau,t')\\
        -\gamma_0^2\int_{0}^{t}dt'\ee^{-\lambda(\tau-t')}\Xi_{\rm TL}(t,t').
\end{multline}
If we write 
\begin{equation}\label{ansatz0}
    \Xi_{\rm TL}(t,t')=\ee^{-\lambda t}\xi(t,t'),
\end{equation}
the previous equation simplifies to
\begin{multline}\label{integro-diff-previous}
		\partial_t{\xi}(\tau,t)=-\ii\dfrac{\gamma_0^2}{\sqrt{2}}\ee^{\lambda t}c_{20}(t)\\
        -\gamma_0^2\int_{0}^{t}dt'\ee^{-\lambda(t-t')}[\xi(\tau,t')+\xi(t,t')].
\end{multline}
Taking here a derivative with respect to $\tau$, we obtain
\begin{equation}
    \partial_{t}[\partial_\tau{\xi}(\tau,t)]=-\gamma_0^2\int_{0}^{t}dt'\ee^{-\lambda(t-t')}\partial_\tau{\xi}(\tau,t').
\end{equation}
This is the same integro-differential equation as \eqref{integrodiff-solvable} for quantity $\partial_\tau{\xi}(\tau,t)$. Its solution is therefore given by \eqref{c10},
\begin{equation}\label{partialxi}
    \partial_\tau{\xi}(\tau,t)=\partial_\tau{\xi}(\tau,0)\upsilon(t).
\end{equation}
Since we are considering the field is initially in the vacuum state, \(c_{1\bm{k}}(0)=0\).  It then follows from \eqref{Pi} that $\Xi(t,0) = 0$ and $\partial_t \Xi(t,0) = 0$ for any quantization volume $V$. In the thermodynamic limit, this implies $\xi(t,0) = 0$ and $\partial_t \xi(t,0) = 0$. As a consequence, Eq.~\eqref{partialxi} yields $\partial_\tau \xi(\tau,t) = 0$, so that $\xi(t,t') = \xi(t')$. We thus arrive at the important result that the two-time dependence of $\Xi_{\rm TL}(t,t')$ factorizes,
\begin{equation}\label{ansatz}
    \Xi_{\rm TL}(t,t')=\ee^{-\lambda t}\xi(t'),
\end{equation}
where $\xi(t')$ satisfies \eqref{integro-diff-previous}, namely
\begin{equation}\label{integrodiffbeta}
    \dot{\xi}(t)=-\ii\frac{\gamma_0^2}{\sqrt{2}}\ee^{\lambda t}c_{20}(t)-2\gamma_0^2\int_{0}^{t}dt'\ee^{-\lambda(t-t')}\xi(t'),
\end{equation}
with the initial condition $\xi(0)=0$. This equation can be solved by means of a Laplace transform, yielding
\begin{equation}
    \xi(t)=-\ii\dfrac{\gamma_0^2}{\sqrt{2}}\int_0^{t}dt'\phi(t-t')\ee^{\lambda t'}c_{20}(t'),
\end{equation}
and hence
\begin{equation}\label{solPiFormal}
    \Xi_{\rm TL}(t,t')=-\ii\frac{\gamma_0^2}{\sqrt{2}}\int_0^{t'}dt''\phi(t'-t'')\ee^{-\lambda (t-t'')}c_{20}(t'').
\end{equation}
Finally, from \eqref{ansatz} we obtain the following result, which plays a central role in solving the two-atom dynamics:
\begin{multline}\label{sumg0}
    \sum_{\bm{k}}g(\omega_{k})\ee^{-\ii\Delta_{k}t}w_k(t')\\=2\int_{0}^{t'}dt'\left[f(t'-t'')\Xi(t,t'')-f(t-t'')\Xi(t',t'')\right]\\
    \xrightarrow{\rm TL} \gamma_0^2\int_{0}^{t'}dt'\left[\ee^{-\lambda t'}\ee^{-\lambda t}-\ee^{-\lambda t}\ee^{-\lambda t'}\right]\ee^{\lambda t''}\xi(t'')\\=0.
\end{multline}

Returning to the evaluation of \eqref{c110step2} in the thermodynamic limit, Eq.~\eqref{sumg0} leads to a significant simplification:
\begin{equation}\label{integrodiff}
    \dot{c}_{20}(t)=-\gamma_0^2\int_{0}^{t}dt'\phi(t-t')\ee^{-\lambda(t-t')}c_{20}(t').
\end{equation}
Differentiating this equation twice with respect to time and using \eqref{phi(t)}, we obtain the differential equation
\begin{equation}\label{diffeq}
    \dddot{c}{\hspace{-0.65ex}}_{20}(t)+3\lambda\ddot{c}_{20}(t)+\left(2\lambda^2+3\gamma_{0}^2\right)\dot{c}_{20}(t)+2\lambda\gamma_{0}^2c_{20}(t)=0,
\end{equation}
subject to the initial conditions
\begin{equation}\label{initialconditions}
    \begin{cases}
        c_{20}(0)=c_{20}(0),\\
        \dot{c}_{20}(0)=0,\\
        \ddot{c}_{20}(0)=-\gamma_0^2\,c_{20}(0).
    \end{cases}
\end{equation}
This equation can be readily solved using standard methods, yielding
\begin{equation}\label{c20}
    c_{20}(t)=c_{20}(0)\zeta(t)=c_{20}(0)\sum_{i=1}^{3}\dfrac{\left(\prod_{j\neq i}z_{j}\right)-\gamma_0^2}{\prod_{j\neq i}(z_i-z_j)}\ee^{z_it},
\end{equation}
where the last equality holds when the characteristic polynomial
\begin{equation}
    P(z) = z^3 + 3 \lambda z^2 + (2 \lambda^2 + 3 \gamma_0^2) z + 2 \lambda \gamma_0^2
\end{equation}
has three distinct roots $z_i$. This is always the case except when
\begin{equation}
    \lambda^2 = \frac{3}{2}
    \left(2 + \sqrt[3]{3 - 2 \sqrt{2}} + \sqrt[3]{3 + 2 \sqrt{2}}\right) \gamma_0^2,
\end{equation}
for which two real roots coincide. We note that $\zeta(t)$ is a real function and is independent of the initial conditions.

Moreover, by comparing \eqref{solPiFormal} with \eqref{integrodiff}, we obtain
\begin{equation}\label{PiTh}
    \Xi_{\rm TL}(t,t')=\frac{\ii}{\sqrt{2}}\ee^{-\lambda(t-t')}\dot{c}_{20}(t'),
\end{equation}
which can be explicitly evaluated using \eqref{c20}.

In order to determine the coefficient $c_{1\bm{k}}(t)$, we substitute \eqref{aux_w} into \eqref{finalsystem2}, obtaining
\begin{multline}
    \!\!\!\!\ii\dot{c}_{1\bm{k}}(t)=\sqrt{2}g(\omega_k)\ee^{\ii\Delta_k t}c_{20}(t)-2\ii\int_{0}^{t}dt'f(t-t')c_{1\bm{k}}(t')\\-2\ii\int_{0}^{t}dt'g(\omega_k)\ee^{\ii\Delta_k t'}\Xi(t,t').
\end{multline}
This equation can be formally integrated using a Laplace transform, leading to
\begin{multline}\label{c1k}
    c_{1\bm{k}}(t)=-\ii\sqrt{2}g(\omega_k)\int_0^{t}dt'\chi(t-t')\ee^{-\ii\Delta_k t'}c_{20}(t')\\
    -2\int_0^{t}dt'\chi(t-t')\int_{0}^{t'}dt''g(\omega_k)\ee^{\ii\Delta_k t''}\Xi(t',t''),
\end{multline}
where $\chi(t):=\mathfrak{L}^{-1}\left[\frac{1}{s+2F(s)}\right](t)$, which in the thermodynamic limit becomes
\begin{multline}\label{upsilon}
    \chi(t)\xrightarrow{\rm TL} \mathfrak{L}^{-1}\left[\frac{s+\lambda}{s(s+\lambda)+\gamma_0^2}\right](t)\\=\text{e}^{-\frac{\lambda t}{2}}\left[\cosh\left(\dfrac{\Omega t}{2}\right)+\dfrac{\lambda}{\Omega}\sinh\left(\dfrac{\Omega t}{2}\right)\right]=\upsilon(t).
\end{multline}

Analogous expressions can be derived for the two-photon coefficients $c_{02\bm{k}}(t)$ and $c_{0\bm{k}\bm{k}'}(t)$. However, as shown in the next section, these quantities are not required to determine the reduced dynamics of the two atoms in the thermodynamic limit.

\subsection{Atomic reduced dynamics and exact master equation} 
We may now calculate the reduced density matrix of the two atoms. Taking the partial trace of \eqref{state2atoms} over the bath degrees of freedom yields
\begin{align}\label{densitymatrix}
    \rho_{S}(t)&=\mathrm{Tr}_{\rm B}\big[\ketbra{\Psi(t)}{\Psi(t)}\big]=\lvert c_{20}(t)\rvert^2\ket{2}\bra{2}\nonumber \\
    &+\left[\lvert c_{10}(t)\rvert^2+Y_2(t)\right]\ketbra{1}{1}+\lvert c_{-,0}\rvert^2\ketbra{-}{-} \nonumber \\
    &+\left[1\!-\!\lvert c_{20}(t)\rvert^2\!-\!\lvert c_{10}(t)\rvert^2\!-\!Y_2(t)\!-\!\lvert c_{-,0}\rvert^2\right]\ketbra{0}{0}\nonumber\\
    &+\big\{ c_{20}(t)c_{10}^{*}(t)\ketbra{2}{1} +c_{20}(t)c_{0}^{*}\ket{2}\bra{0}\nonumber \\
    &\qquad+c_{20}(t)c_{-,0}^{*}\ketbra{2}{-}+\left[c_{10}(t)c_{0}^{*}+Y_{1}^{*}(t)\right]\ketbra{1}{0}\nonumber \\
    &\qquad+c_{10}(t)c_{-,0}^{*}\ketbra{1}{-}+c_{0}c_{-,0}^{*}\ketbra{0}{-}+\mathrm{h.c.}\big\},
\end{align}	
where
\begin{align}
        &Y_{1}(t):=\sum_{\bm{k}}c_{0\bm{k}}(t)c_{1\bm{k}}^{*}(t),\label{ints1}\\
        &Y_{2}(t):=\sum_{\bm{k}}\lvert c_{1\bm{k}}(t)|^2,\label{ints2}
\end{align}
and we have used the condition $\mathrm{Tr}[\rho_S(t)]=1$ to remove the explicit dependence on the two-photon coefficients $c_{02\bm{k}}(t)$ and $c_{0\bm{k}\bm{k}'}(t)$.

We now evaluate $Y_{1,2}(t)$ in the thermodynamic limit. From \eqref{ints1} and the formal expressions \eqref{c0k} and \eqref{c1k}, we obtain
\begin{multline}
    \!\!\!\!Y_{1}(t)=2\!\int_0^{t}\!dt'c_{10}(t')\int_0^{t}\!dt''\chi^{*}(t-t'')f(t'-t'')c_{20}^{*}(t'')\\
    +2\ii\sqrt{2}\int_0^tdt'c_{10}(t')\int_0^{t}dt''\chi^{*}(t-t'')\\
    \times\int_0^{t''}d t''' f(t'-t''')\Xi^{*}(t'',t'''),
\end{multline}
where we have also made use used \eqref{corrfunction}. Taking the thermodynamic limit $V\to\infty$, and invoking \eqref{corrfunctionth}, \eqref{PiTh}, and \eqref{upsilon}, this expression becomes
\begin{align}\label{Y1TL}
    Y_{1}(t)\!\xrightarrow{\rm TL}\mbox{}& \gamma_0^2\!\int_0^{t}\!\!dt'c_{10}(t')\!\!\int_0^{t}\!\!dt''\upsilon(t\!-\!t'')\ee^{-\lambda|t'-t''|}c_{20}^{*}(t'')\nonumber \\
    &\mbox{}\!\!\!\!+\gamma_0^2\int_0^tdt'c_{10}(t')\int_0^{t}dt''\upsilon(t-t'')\nonumber \\
    &\ \mbox{}\!\! \times\!\int_0^{t''}\!dt'''\ee^{-\lambda|t'-t'''|}\ee^{-\lambda(t''-t''')}\dot{c}_{20}^{*}(t''').
\end{align}
The nested integrals can be simplified by applying a Laplace transform with respect to the variable $t$:
\begin{align}
\mathfrak{L}&\left[{\textstyle \int_0^{t}dt'\upsilon(t-t')\int_0^{t'}dt''\ee^{-\lambda|\tau-t''|}\ee^{-\lambda(t'-t'')}\dot{c}_{20}^{*}(t'')}\right](s)\nonumber\\
    &=\{\mathfrak{L}[\upsilon(t)](s)\} \{\mathfrak{L}[\ee^{-\lambda t}](s)\} \{\mathfrak{L}[\ee^{-\lambda|\tau-t|}\dot{c}_{20}^{*}(t)](s)\}  \nonumber\\
    &=\frac{1}{s(s+\lambda)+\gamma_0^2}\mathfrak{L}[\ee^{-\lambda|\tau-t|}\dot{c}_{20}^{*}(t)](s).
\end{align}
By now taking the inverse Laplace transform, we obtain the identity
\begin{multline}\label{intsidentity}
    \int_0^{t}dt'\upsilon(t-t')\int_0^{t'}dt''\ee^{-\lambda|\tau-t''|}\ee^{-\lambda(t'-t'')}\dot{c}_{20}^{*}(t'')\\
    =\int_0^{t}dt'\eta(t-t')\ee^{-\lambda|\tau-t'|}\dot{c}_{20}^{*}(t'),
\end{multline}
where
\begin{equation}\label{xi}
    \eta(t)=\mathfrak{L}^{-1}\left[\dfrac{1}{s(s+\lambda)+\gamma_0^2}\right](t)=\dfrac{2}{\Omega}\ee^{-\frac{\lambda t}{2}}\sinh\left(\dfrac{\Omega t}{2}\right).
\end{equation}
Equation \eqref{Y1TL} can therefore be rewritten as
\begin{align}\label{Y1}
    Y_{1}(t)\xrightarrow{\rm TL}\gamma_0^2\int_0^{t}& dt' c_{10}(t')\int_0^{t}dt''\ee^{-\lambda|t'-t''|}\nonumber\\
    &\!\!\!\times[\upsilon(t-t'')c_{20}^{*}(t'')+\eta(t-t'')\dot{c}_{20}^{*}(t'')].
\end{align}
Defining $Y_1(t):=c_{10}(0)c_{20}^{*}(0)I_1(t)$ and using \eqref{c10} and \eqref{c20}, one may equivalently write
\begin{align}
    I_{1}(t)\xrightarrow{\rm TL}\gamma_0^2\int_0^{t}& dt' \upsilon(t')\int_0^{t}dt''\ee^{-\lambda|t'-t''|}\nonumber\\
    &\times[\upsilon(t-t'')\zeta(t'')+\eta(t-t'')\dot{\zeta}(t'')],
\end{align}
from which it is clear that $I_1(t)$ is a real function independent of the initial conditions. Note that to evaluate these integrals, the absolute value may be removed using the standard decomposition
\begin{equation}\label{integraltrick}              
\int_{0}^{t}dt''\int_{0}^{t}dt'=\int_{0}^{t}dt''\int_{0}^{t''}dt'+\int_{0}^{t}dt'\int_{0}^{t'}dt''.
\end{equation}
The resulting integrals are elementary, involving only linear combinations of exponential functions. The final expressions can therefore be obtained in closed form, but are not particularly compact and are omitted for brevity.

A similar procedure gives
\begin{align}\label{Y2}
    Y_{2}&(t)\xrightarrow{\rm TL}\gamma_0^2\int_0^{t}\!\!dt' [\upsilon(t\!-\!t')c_{20}(t')+\eta(t\!-\!t')\dot{c}_{20}(t')]\nonumber\\
    &\!\times\!\!\int_0^{t}\!\!dt''\ee^{-\lambda|t'-t''|}[\upsilon(t\!-\!t'')c_{20}^{*}(t'')+\eta(t\!-\!t'')\dot{c}_{20}^{*}(t'')],
\end{align}
which, upon defining $Y_2(t):=|c_{20}(0)|^2 I_2(t)$, can be written as
\begin{align}\label{I2}
    I_{2}&(t)\xrightarrow{\rm TL}\gamma_0^2\int_0^{t}\!\! dt' [\upsilon(t-t')\zeta(t')+\eta(t-t')\dot{\zeta}(t')]\nonumber\\
    &\times\!\!\int_0^{t}\!\!dt''\ee^{-\lambda|t'-t''|}[\upsilon(t-t'')\zeta(t'')+\eta(t-t'')\dot{\zeta}(t'')].
\end{align}
Again, $I_2(t)$ is a real function independent of the initial conditions and can be expressed in closed form using \eqref{integraltrick}.

Thus, after taking the thermodynamic limit, all time-dependent coefficients appearing in \eqref{densitymatrix} are fully determined by \eqref{c10}, \eqref{c20}, \eqref{Y1}, and \eqref{Y2}. The reduced dynamics is therefore completely solved. 

To further characterize its structure, it is useful to derive the master equation governing the dynamics. Taking the time derivative of \eqref{densitymatrix} and expressing the result in the Dicke basis, one finds by direct substitution that
\begin{equation}\label{masterequation_app}
    \dfrac{d\rho_S(t)}{dt}=\!\!\!\sum_{m,n=1}^{3}\!\!\!\Gamma_{mn}(t)\!\!\left[L_{m}\rho_S(t)L_{n}^{\dagger}\!-\!\tfrac{1}{2}\!\left\{L_{n}^{\dagger}L_{m},\rho_S(t)\right\}\right]\!,
\end{equation}
with the jump operators
\begin{equation}
    \begin{cases}
        L_1=\ket{1}\bra{2},\\
        L_2=\ket{0}\bra{1},\\
        L_3=\ket{0}\bra{2},
    \end{cases}
\end{equation}
and the nonvanishing coefficients $\Gamma_{mn}(t)$ given by
\begin{align}
        \Gamma_{11}(t)&=-\frac{1}{\zeta^{2}(t)}\left[2\frac{\dot{\upsilon}(t)}{\upsilon(t)}I_2(t)-\dot{I}_2(t)\right],\label{Gamma11}\\
        \Gamma_{22}(t)&=-2\frac{\dot{\upsilon}(t)}{\upsilon(t)},\\
        \Gamma_{33}(t)&=\frac{1}{\zeta^{2}(t)}\left[2\frac{\dot{\upsilon}(t)}{\upsilon(t)}I_2(t)-\dot{I}_2(t)\right]-2\frac{\dot{\zeta}(t)}{\zeta(t)},\\
        \Gamma_{12}(t)&=\Gamma_{21}(t)=\frac{1}{\zeta(t)\upsilon(t)}\left[\dot{I}_1(t)-\frac{\dot{\upsilon}(t)}{\upsilon(t)}I_1(t)\right].\label{Gamma12}
\end{align}
All coefficients $\Gamma_{mn}(t)$ are therefore real and independent of the initial conditions, confirming that \eqref{masterequation_app} is indeed the master equation governing the reduced dynamics.

\section{Eternal non-Markovianity for \texorpdfstring{$N=2$}{N=2} in the limit \texorpdfstring{$\lambda\gg\gamma_0$}{λ ≫ γ₀}}\label{app:B}
\setcounter{equation}{0}
From Eqs. \eqref{sol1exc}, \eqref{c20}, \eqref{xi}, and \eqref{I2}, it can be easily verified that $\zeta(t)$, $\upsilon(t)$, and $I_{2}(t)$ can all be written as linear combinations of exponential functions. The corresponding exponents can be expressed in terms of $\omega_{1,2}:=-\tfrac{1}{2}(\lambda\pm\Omega)$ and $z_{j}$. To simplify the notation, let us define
\begin{figure*}[t]
	\includegraphics[width=\textwidth]{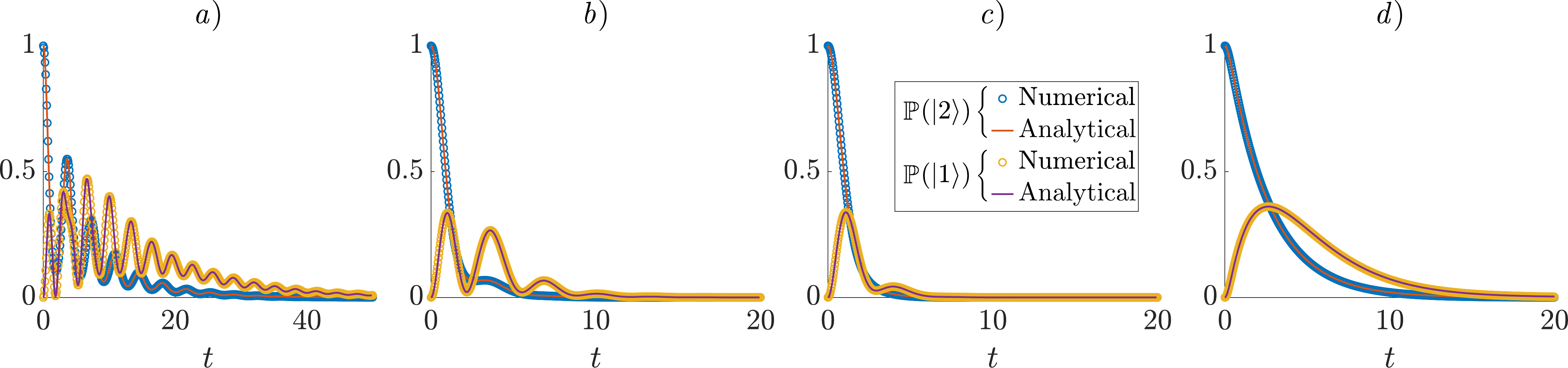}
	\caption{ Comparison of the analytical (lines) and numerical (circles) time evolution of the populations \eqref{populations} for $\textit{a}$) $\lambda=0.1$, $\textit{b})$ $\lambda=1/2$,  $\textit{c})$ $\lambda=1$ and $\textit{d})$ $\lambda=5$, for the case $N=2$. The value $\omega_0 = 10^3$ is chosen. All quantities are in units of $\gamma_0=1$.}
	\label{Fig:9}
\end{figure*}
\begin{equation}
    y(t)=\sum_{n} C_{n}\ee^{\nu_{n}t}=\sum_{n} C_{n}\ee^{\bar{\nu}_{n}\tau}=:\bar{y}(\tau),
\end{equation}
where $\lambda$ is the width of the spectral density \eqref{spectraldensity}, and we have introduced the dimensionless quantities $\bar{\nu}_{n}=\nu_{n}/\lambda$ and $\tau=\lambda t\geq0$. With this notation, the canonical decay rate $\gamma_{3}(t)$ can be rewritten as
\begin{align}
    \gamma_3(t)&=\frac{1}{\zeta^{2}(t)}\left[2\frac{\dot{\upsilon}(t)}{\upsilon(t)}I_2(t)-\dot{I}_2(t)\right]-2\frac{\dot{\zeta}(t)}{\zeta(t)}\nonumber\\
    &=\lambda\left\{\frac{1}{\bar{\zeta}^{2}(\tau)}\left[2\frac{\dot{\bar{\upsilon}}(t)}{\bar{\upsilon}(t)}\bar{I}_2(\tau)-\dot{\bar{I}}_2(\tau)\right]-2\frac{\dot{\bar{\zeta}}(\tau)}{\bar{\zeta}(\tau)}\right\}\nonumber \\
    &=:\bar{\gamma}_3(\tau).
\end{align}
By expanding all coefficients and exponents in powers of the dimensionless parameter $\gamma_0/\lambda$, one obtains
\begin{multline}
    \bar{\gamma}_{3}(\tau)=\dfrac{\gamma_0^6}{\lambda^5}\ee^{-3\tau}\!\left\{1+\ee^{\tau}\!\left[2\tau\!-\!5\!+\!\ee^{\tau}\!\left(4\tau^2\!-2\tau\!+\!7\right)\right]\right\}\\-3\dfrac{\gamma_0^6}{\lambda^5}+O\left(\dfrac{\gamma_0^8}{\lambda^7}\right).
\end{multline}
Therefore, taking the limit $\gamma_0/\lambda\rightarrow0$, we find
\begin{multline}\label{lim-def-G}
    G(\tau):=\lim_{\gamma_0/\lambda\rightarrow0}\dfrac{\lambda^5}{3\gamma_0^6}\bar{\gamma}_{3}(\tau)\\
    =\dfrac{\ee^{-3\tau}}{3}\left\{1+\ee^{\tau}\left[2\tau-5+\ee^{\tau}\left(4\tau^2-2\tau+7\right)\right]\right\}-1.
\end{multline}
It is straightforward to verify that $G(0)=0$ and $\lim_{\tau\rightarrow\infty}G(\tau)=-1$. To analyze the sign of $G(\tau)$, let us define $H(\tau)=\ee^{3\tau}G(\tau)$. A direct calculation yields
\begin{equation}\label{initcondsH}
    H(0)=H'(0)=H''(0)=H'''(0)=H^{(\rm iv)}(0)=0,
\end{equation}
and
\begin{multline}
    H^{(\rm v)}(\tau)=-243\ee^{3\tau}\\+\dfrac{\ee^{\tau}}{3}\left[2\tau+5+64\ee^{\tau}\left(2\tau^2+9\tau+11\right)\right].
\end{multline}
Hence, $H^{(\mathrm{v})}(0)=-20/3$. Introducing $\tilde{H}^{(\mathrm{v})}(\tau):=\ee^{-3\tau}H^{(\mathrm{v})}(\tau)$, we find $\tilde{H}^{(\mathrm{v})}(0)=-20/3$, and
\begin{multline}
    \dfrac{d}{d\tau}\tilde{H}^{(\rm v)}(\tau)=-\dfrac{4}{3}\ee^{-2\tau}(\tau+2)\\-\dfrac{64\ee^{-\tau}}{3}(2\tau^2+5\tau+2)<0.
\end{multline}
Therefore, $\tilde{H}^{(\rm v)}(\tau)$ is a strictly decreasing function, and since $\tilde{H}^{(\rm v)}(0)<0$, it follows that $\tilde{H}^{(\rm v)}(\tau)<0$. Consequently, $H^{(\rm v)}(\tau)=\ee^{3\tau}\tilde{H}^{(\rm v)}(\tau)<0$. 
Successive integration, together with \eqref{initcondsH}, then implies that $H(\tau)<0$ for all $\tau>0$, and so
\begin{equation}
    G(\tau)=\ee^{-3\tau}H(\tau)<0\quad\forall t>0.
\end{equation}
It follows from \eqref{lim-def-G} that, in the limit $\gamma_0/\lambda\to0$, the canonical decay rate $\gamma_3(t)$ approaches a negative function that vanishes only at $t=0$. Since a negative canonical decay rate implies non-Markovianity, the dynamics is non-Markovian for all $t>0$ in this limit—a property known as eternal non-Markovianity.

\section{Cross-check between the analytical and pseudomode solutions}
\setcounter{equation}{0}
In order to further validate the pseudomode treatment employed throughout the manuscript, in this appendix we explicitly compare the numerical pseudomode dynamics with the exact analytical solution obtained for the case $N=2$ in the previous section. Fig. \ref{Fig:9} shows the time evolution of the atomic Dicke populations 
\begin{equation}\label{populations}
    \mathbb{P}(\ket{n})=\bra{n}\rho_S(t)\ket{n},
\end{equation}
obtained from both approaches ([only two of the three populations are shown, since the third is determined from the other two with the condition ${\rm Tr}(\rho_S)=1$]. The solid lines correspond to the analytical solution, while the circles correspond to the pseudomode simulation. The curves overlap within numerical precision over the entire evolution. For the particular initial condition considered here, namely both atoms initially prepared in the excited state, the reduced atomic dynamics remains incoherent, i.e., $\bra{m}\rho_{S}\ket{n}=0$ if $m\neq n$ throughout the evolution. Consequently, the reduced density matrix is completely determined by the state populations alone. Therefore, the agreement observed in the populations implies agreement of the full reduced atomic density matrix between the analytical and numerical treatments.

\end{document}